%% file: main.tex
\documentclass[a4paper, 12pt]{article}
\usepackage{amsmath}
\usepackage{amssymb}
\usepackage{amsfonts}
\usepackage{array}
\usepackage{tikz}
\usetikzlibrary{calc}
\usepackage{pgfplots}
\pgfplotsset{compat=1.11}
\usepackage{subcaption}
\usepackage{mathrsfs}
\usepackage{multirow}
\usepackage{siunitx}
\usepackage{microtype}
\usepackage{changes}
\newcommand{\bea}{\begin{eqnarray*}}
\newcommand{\eea}{\end{eqnarray*}}
\newcommand{\beao}{\begin{eqnarray}}
\newcommand{\eeao}{\end{eqnarray}}

\newcommand{\RR}{{\mathbb R}}

\newcommand{\mev}{\mega\electronvolt}
\newcommand{\gev}{\giga\electronvolt}
\newcommand{\tev}{\tera\electronvolt}

\usepackage{lineno}
\usepackage[hidelinks]{hyperref}

\usepackage{xspace}

\usepackage{booktabs}
\usepackage{listings}
\usepackage{float}
\floatstyle{plaintop}
\newfloat{lstfloat}{tbp}{lop}
\floatname{lstfloat}{Listing}
\input{lstinit.tex}

\newcommand{\tomopt}{\textsc{TomOpt}\xspace}

\newcommand{\pytorch}{\textsc{PyTorch}\xspace}
\newcommand{\geant}{\textsc{Geant}4\xspace}

\usepackage{xargs}

\usepackage{authblk}

\newcommand{\eg}{e.g.\xspace}
\newcommand{\ie}{i.e.\xspace}
\begin{document}

\lstset{
            tabsize=2,
            lineskip=-1pt,
            rulecolor=,
            columns=fullflexible,
            upquote=true,
            aboveskip={0.pt},
            columns=fixed,
            showstringspaces=false,
            extendedchars=true,
            breaklines=false,
            prebreak = \raisebox{0ex}[0ex][0ex]{\ensuremath{\hookleftarrow}},
            escapechar=@,
            frame=no,
            showtabs=false,
            showspaces=false,
            showstringspaces=false,
            identifierstyle=\ttfamily,
            keywordstyle=\color[rgb]{1.0,0,0}\ttfamily,
            keywordstyle=[1]\color[rgb]{0,0,0.75}\ttfamily,
            keywordstyle=[2]\color[rgb]{0.5,0.0,0.0}\ttfamily,
            keywordstyle=[3]\color[rgb]{0.127,0.427,0.514}\ttfamily,
            keywordstyle=[4]\color[rgb]{0.4,0.4,0.4}\ttfamily,
            commentstyle=\color[rgb]{0.133,0.545,0.133}\ttfamily\itshape,
            stringstyle=\color[rgb]{0.639,0.082,0.082}\ttfamily,
        }


\title{Progress in End-to-End Optimization\\ of Detectors for Fundamental Physics \\with Differentiable Programming}

\author[1,2]{Max~Aehle}
\author[3,4]{Lorenzo~Arsini}
\author[5]{R.~Bel\'en~Barreiro}
\author[6]{Anastasios~Belias}
\author[7]{Florian~Bury}
\author[8]{Susana~Cebrian}
\author[9]{Alexander~Demin}
\author[10]{Jennet~Dickinson}
\author[1,11,12]{Julien~Donini}
\author[1,12,13]{Tommaso~Dorigo\footnote{dorigo@pd.infn.it}}
\author[13]{Michele~Doro}
\author[1,2]{Nicolas~R.~Gauger}
\author[1,14]{Andrea~Giammanco\footnote{andrea.giammanco@uclouvain.be}}
\author[10]{Lindsey~Gray}
\author[15,16]{Borja~S.~González}
\author[17]{Verena~Kain}
\author[1,18]{Jan~Kieseler}
\author[1,2]{Lisa~Kusch}
\author[19]{Marcus~Liwicki}
\author[20]{Gernot~Maier}
\author[1,11,21]{Federico~Nardi}
\author[1,9]{Fedor~Ratnikov}
\author[22]{Ryan~Roussel}
\author[23]{Roberto~Ruiz~de~Austri}
\author[19]{Fredrik~Sandin}
\author[17]{Michael Schenk}
\author[21]{Bruno~Scarpa}
\author[17]{Pedro~Silva}
\author[1,13]{Giles~C.~Strong}
\author[1,24]{Pietro~Vischia\footnote{pietro.vischia@cern.ch}}

\affil[1]{MODE Collaboration, https://mode-collaboration.github.io/}
\affil[2]{Chair for Scientific Computing, University of Kaiserslautern-Landau, Germany}
\affil[3]{La Sapienza Università di Roma, Rome, Italy}
\affil[4]{Istituto Nazionale di Fisica Nucleare, Sezione di Roma, Italy}
\affil[21]{Instituto de Física de Cantabria, UC-CSIC, Spain}
\affil[6]{GSI, Germany}
\affil[7]{University of Brisol, United Kingdom}
\affil[8]{Centro de Astropart\'iculas y F\'isica de Altas Energ\'ias (CAPA), Universidad de Zaragoza, Spain}
\affil[9]{HSE University, Russia}
\affil[10]{Fermi National Accelerator Laboratory, USA}
\affil[11]{Universit\'e Clermont Auvergne, Laboratoire de Physique de Clermont, CNRS/IN2P3, France}
\affil[12]{Universal Scientific Education and Research Network}
\affil[13]{Istituto Nazionale di Fisica Nucleare, Sezione di Padova, Italy}
\affil[14]{Centre for Cosmology, Particle Physics and Phenomenology (CP3), Universit\'e catholique de Louvain, Belgium}
\affil[15]{Laborat\'orio de Instrumenta\c{c}\~ao e F\'\i{}sica Experimental de Part\'\i{}culas -- LIP, Lisbon, Portugal}
\affil[16]{Instituto Superior T\'ecnico -- IST, Universidade de Lisboa -- UL, Lisbon, Portugal}
\affil[17]{CERN, Switzerland}
\affil[18]{Karlsruhe Institute of Technology, Germany}
\affil[19]{Lule{\aa} University of Technology, Sweden}
\affil[20]{Deutsches Elektronen-Synchrotron (DESY), Germany}
\affil[21]{Universit\'a degli Studi di Padova, Italy}
\affil[22]{Stanford Linear Accelerator Center, United States}
\affil[23]{Instituto de Física Corpuscular, UV-CSIC, Spain}
\affil[24]{Universidad de Oviedo and ICTEA, Spain}

\maketitle

\clearpage

\begin{abstract}
In this article we examine recent developments in the research area concerning the creation of end-to-end models for the complete optimization of measuring instruments. The models we consider rely on differentiable programming methods and on the specification of a software pipeline including all factors impacting performance---from the data-generating processes to their reconstruction and the extraction of inference on the parameters of interest of a measuring instrument---along with the careful specification of a utility function well aligned with the end goals of the experiment. 

Building on previous studies originated within the MODE Collaboration, we focus specifically on applications involving instruments for particle physics experimentation, as well as industrial and medical applications that share the detection of radiation as their data-generating mechanism.
\end{abstract}

\tableofcontents
\clearpage

\section{Introduction}
\label{sec:intro}

In the course of the past few centuries, progress in fundamental science has tracked quite closely the corresponding progress in our technological skills and the capability of conceiving, constructing, and operating more and more complex measuring instruments. An example of such synergy is offered by the history of particle physics, whose research in the past seventy years posed increasingly challenging demands on the performance of devices that base their functioning on the interaction of radiation with matter. 

While this field has taken advantage and is still benefiting from continuing progress in the production of more and more performing gaseous and solid-state detectors and related electronics, we believe that the most notable and ground-breaking technological advancement that has taken place over the past two decades is the coming of age of machine learning methods, further enhanced by progressively cheaper and more powerful computing infrastructures. It is therefore only natural to exploit that advancement in the design of instruments meant to further our understanding of Nature. Besides, the sheer scale of the experiments we perform today lends itself naturally as a challenge to be addressed with solutions offered by computer science. Specifically, the non-trivial interrelation between the outputs of the large number of components and subsystems making up a modern particle detector causes a significant difference between the result of optimization procedures considering each the subsystems as separate entities, which are necessarily based on local figures of merit, and the global optimization of the system as a whole, which can directly be based on an utility function aligned to the experimental goals. An example of this sort of misalignment is well known to data acquisition specialists at hadron colliders: the fixed nature of the total bandwidth available for data collection poses luminosity-dependent constraints on the effective cross section of data selection recipes corresponding to each individual trigger stream, each of which results from careful optimization of selection strategies performed separately from one another. The occasional urgency of reducing dead time that occurs e.g. during a higher-than-average luminosity run---when the average accept rate is too high, which induces a finite probability that accepted events fail to get stored---forces rate reduction strategies that cannot account for the overall scientific goals of the experiment.

By and large, the engine under the hood of most modern machine learning methods is what has come to be called \emph{Differentiable Programming} (DP). DP relies on automatic differentiation (AD) procedures, which are nowadays offered by several common tools (\pytorch~\cite{pytorch}, TensorFlow~\cite{abadi2016tensorflow}, and JAX~\cite{jax2018github}, among others). Benefiting from the automatic computation of derivatives of whole pieces of software simplifies greatly the search for the extremum of arbitrarily complex utility functions, by employing the standard technique of gradient descent. While DP is not the only solution to large-scale holistic optimization problems, and may not be viable in specific cases, we find it particularly suitable to allow for a unified modeling of the various parts of a global optimization task of the kinds of interest in our research area. 

The intrinsically stochastic nature of the data-generating processes of interaction of radiation with matter, arising from quantum phenomena, is a source of additional, conspicuous complications in the way of the creation of a full differentiable model of the whole problem. Workarounds based on the creation of surrogate models often provide viable solutions, which are however invariably rather specific to the problem at hand. This is one of the main stumbling blocks in the creation of versatile, multipurpose architectures for end-to-end optimization. Still, we believe that the solution of a significant number of different problems of low to medium complexity will empower our community to construct solutions to still harder, larger-scale experiment optimization tasks, such as those concerning detectors for a future very-high-energy particle collider. 

The present document constitutes an update and an extension of the ideas and the use cases that some of us recently described in a previous work~\cite{whitepaper}. We have structured it as follows.
In Section~\ref{sec:cs} we discuss the state of the art and the recent developments of computer science tools involved in the deployment of end-to-end models for optimization. The following sections discuss separate use cases for end-to-end optimization. In Section~\ref{sec:tomopt} we discuss applications to muon tomography, where we have made our first attempts at a full solution of the optimization problem. In Section~\ref{sec:calopt} we consider the possible advancements in high-precision calorimetry by exploiting differentiable models to study the integration and hybridization of tracking and calorimetry devices. Section~\ref{sec:accel} discusses applications in accelerator optimization. Section~\ref{sec:astro} discusses several use cases from fundamental research in experimental astroparticle physics. We discuss in Section~\ref{sec:neuro} the possibility of integrating neuromorphic computing devices in particle detectors and of exploiting this innovative computing paradigm for the optimization tasks at the focus of this work.
Section~\ref{sec:medical} deals with progress in optimization tasks for the benefit of detectors for medical applications. Finally, we provide a brief overlook of this young but promising new field of research in Section~\ref{sec:concl}.

\clearpage
\section {Progress in AD methods}
\label{sec:cs}

\subsection{Towards Algorithmic Differentiation of GATE/\geant}\label{sec:ad-of-geant4}

Simulations of the detection process are an essential step in the assessment of a proposed detector design.
Often, complex Monte Carlo simulators like \geant~\cite{GEANT,allison_geant4_2006,allison_recent_2016} are employed for this task, as they provide the most realistic and adaptable computational models for the interactions between particles and the detector. In addition, however, many simplified simulators and surrogate models have been proposed in the literature.

Naturally, detector optimization toolchains derived from assessment pipelines contain at least one of these simulators or models. If changes of the detector design have only minor effects on the particles, as \eg in the muon tomography setup of \tomopt (described in Section~\ref{sec:tomopt}), there is no need to differentiate through the particle simulation. Otherwise, \eg in the case that the geometrical layout of absorber material in the detector changes, the particle simulation code becomes part of the objective function to be optimized.

One possible approach to algorithmically differentiate such a complicated objective function would be to replace any Monte Carlo particle simulators by surrogate models. Using such surrogate models has the following advantages:
\begin{itemize}
\item after a training phase, surrogate models can run faster;
\item surrogate models might provide a ``smoother'' approximation that can be better suited for gradient-based optimization;
\item ultimately, applying AD/DP directly to big software projects like \geant is generally considered a severe technical challenge, given their size and complexity.
\end{itemize}

However, the training and use of surrogate models consumes developer and computation time as well. Because of the additional modelling step, assessment pipelines based on a surrogate model may produce less realistic results and thus steer an optimization procedure towards sub-optimal or biased designs. In order to compare the optimization results achievable with surrogate models and the direct application of AD/DP to particle physics Monte Carlo codes, we first need to overcome the technical challenge associated with the latter.

To this end, we recently created the AD tool Derivgrind\footnote{\url{https://github.com/SciCompKL/derivgrind}} \cite{aehle_forward-mode_2022,aehle_reverse-mode_2022}. 
AD tools identify real-arithmetic operations in the \emph{primal program}, and create a program that computes the derivative of  the primal program's \emph{output variables} with respect to its \emph{input variables}, where both sets of variables are defined by the user. AD tools exist in the shape of, \eg, execution environments for domain-specific languages, programs transforming the source code, class definitions, or compiler plugins. 
Unlike traditional source-code-based AD tools,
Derivgrind operates on the machine code of the compiled primal program just before it runs on the processor.
In the best case, machine-code-based AD reduces the developer interaction with the primal program's source code 
to the necessary minimum of inserting macro calls indicating input and output variables.
Derivgrind is therefore well-positioned for initial explorative studies about the direct application 
of AD/DP to complex, cross-language or partially closed-source software projects.

Derivgrind utilizes the dynamic binary instrumentation framework Valgrind~\cite{valgrind-paper} 
to discover floating-point instructions in the compiled primal program, and to insert the corresponding
AD logic. Some real arithmetic can also be performed by binary manipulation of 
floating-point data. Derivgrind handles the most important constructs of this sort correctly.
 However, more obscure ``bit-tricks'' are hard to discover, and accordingly must not be present in the primal program.

\begin{figure}
\centering
\begin{tikzpicture}
\begin{axis}[xtick={229.6,230,230.4}, xticklabels={229.6,230,230.4}, extra x tick style={grid=major}, xmajorgrids, xlabel={beam energy $x$ in \SI{1}{\mega\eV}},xlabel style={align=center,font=\scriptsize},tick label style={font=\scriptsize},ylabel style={align=center,font=\scriptsize}, ylabel={hit coordinate \\ $f_i(x)$ in \si{\milli\meter}},height=4cm,width=0.8\textwidth,legend pos=south west,legend style={font=\scriptsize}]
\addplot[blue,mark=*,mark size=0.5pt,only marks] table[x index=0, y index=2] {figs/ad_for_geant4/around_230/yy};
\addplot[brown,mark=diamond*,mark size=0.5pt,only marks] table[x index=0, y index=4] {figs/ad_for_geant4/around_230/yy};
\legend{first layer ($i=1$),second layer ($i=2$)}
\end{axis}
\end{tikzpicture}
\vspace{-0.3cm}
\caption{Graph of the function that we apply AD to.}
\label{fig:around_230}
\end{figure}
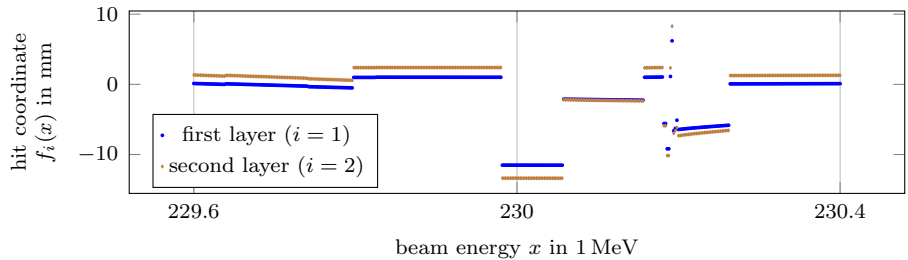

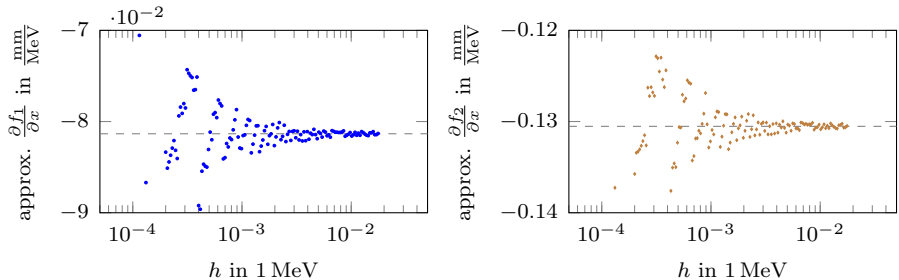
\begin{figure}
\centering
\begin{tikzpicture}
\begin{axis}[xmode=log,align=center,xlabel={$h$ in \SI{1}{\mega\eV}},xlabel style={align=center,font=\scriptsize},tick label style={font=\scriptsize},ylabel style={align=center,font=\scriptsize}, ylabel={approx.\ $\tfrac{\partial f_1}{\partial x}$ in \si[per-mode=fraction,fraction-function=\tfrac]{\milli\meter\per\mega\eV}},height=4cm,width=0.4\textwidth,xmin=0.5e-4,xmax=0.5e-1,ymin=-0.09,ymax=-0.07,restrict y to domain=-1:1,ytick distance=0.01]
\addplot[blue,mark=*,mark size=0.5pt,only marks] table[x index=0, y index=1] {figs/ad_for_geant4/diffquot/diffquot1};
\draw[gray,dashed] (axis cs:0.5e-4,-0.0813391) -- (axis cs:0.5e-1,-0.0813391);
\end{axis}
\end{tikzpicture}
\begin{tikzpicture}
\begin{axis}[xmode=log,align=center,xlabel={$h$ in \SI{1}{\mega\eV}},xlabel style={align=center,font=\scriptsize},tick label style={font=\scriptsize},ylabel style={align=center,font=\scriptsize}, ylabel={approx.\ $\tfrac{\partial f_2}{\partial x}$ in \si[per-mode=fraction,fraction-function=\tfrac]{\milli\meter\per\mega\eV}},height=4cm,width=0.4\textwidth,xmin=0.5e-4,xmax=0.5e-1,ymin=-0.14,ymax=-0.12,restrict y to domain=-1:1,ytick distance=0.01]
\addplot[brown,mark=diamond*,mark size=0.5pt,only marks] table[x index=0, y index=1] {figs/ad_for_geant4/diffquot/diffquot2};
\draw[gray,dashed] (axis cs:0.5e-4,-0.130524) -- (axis cs:0.5e-1,-0.130524);
\end{axis}
\end{tikzpicture}
\vspace{-0.3cm}
\caption{Central difference quotients $\tfrac{f_i(x_0+h)-f_i(x_0-h)}{2h}$ (blue and brown markers) around $x_0=\SI{230}{\mega\eV}$, and the derivative computed with Derivgrind (dashed line) after changing \lstinline|G4Log| to \lstinline|log|.}
\label{fig:diffquot}
\end{figure}

In this work, we apply Derivgrind to GATE (v9.2)~\cite{GATE}, a software package built on top of \geant (v11.0.0) for simulations in medical imaging and radiotherapy.
Our setup is related to a proton computed tomography (pCT) scanning process with a digital tracking calorimeter (DTC) developed by the Bergen pCT collaboration \cite{alme_high-granularity_2020}; in Section~\ref{sec:pct}, we give more details on the tomographic reconstruction algorithm that makes use of the post-processed DTC measurements.
As in Ref.~\cite{aehle-derivatives-2023}, we consider the beam energy as an input variable $x$ for AD, simulate a single proton passing through a human head and the DTC,
and record the first coordinate of the hit in the $i$-th tracking layer of the DTC as the AD output variable $f_i(x)$ for $i=1,2$. As apparent from Figure~\ref{fig:around_230},
these functions $f_1$ and $f_2$ are piecewise differentiable. Figure~\ref{fig:diffquot} shows the central difference quotient $\tfrac{f_i(x_0+h)-f_i(x_0-h)}{2h}$ around $x_0=\SI{230}{\mega\eV}$ for various values of $h$.
Mathematically, this difference quotient converges to the derivative $\tfrac{\partial f_i}{\partial x}(x_0)$ in the limit $h\to 0$. On a computer however, floating-point inaccuracies become large for small $h$.

Listings~\ref{lst:insertion-forward} and~\ref{lst:insertion-reverse} show our insertions into GATE's source code, 
which was slightly refactored beforehand for the purpose of presentation. 
In the first code block of either listing, GATE reads the energy $x$ from the configuration file and sets the respective property of the
beam source object. The second code block is run whenever the proton hits a layer, to assemble GATE's output data. 
The calculations performed inbetween involve \geant, which we do not modify at first.
The inserted macros are defined in a header \lstinline|derivgrind.h| and perform a ``client request'' from the primal program 
to the Derivgrind process running it, basically declaring the AD input and output variables.

\begin{lstfloat}
\caption{Insertions into the source code of GATE for forward-mode differentiation. Seed the input variable (beam energy), and print the dot value of the output variables (hit positions). Additionally, the header \lstinline|derivgrind.h| must be included.}
\label{lst:insertion-forward}
\vspace{0.2cm}
\begin{lstlisting}[language=c,frame=single,basicstyle=\ttfamily\scriptsize,linebackgroundcolor={\ifnum\value{lstnumber}>2\ifnum\value{lstnumber}<5\color{green!20}\fi\fi}]
   if (command == pEnergyCmd) {
     double energy = pEnergyCmd->GetNewDoubleValue(newValue);
+    double one = 1.0;
+    DG_SET_DOTVALUE(&energy,&one,sizeof(double));
     pSourcePencilBeam->SetEnergy(energy);
   }
\end{lstlisting}
\begin{lstlisting}[language=c,frame=single,basicstyle=\ttfamily\scriptsize,linebackgroundcolor={\ifnum\value{lstnumber}>1\ifnum\value{lstnumber}<6\color{green!20}\fi\fi}]
   if (m_rootHitFlag) m_treeHit->Fill();
+  float pos = *(float*)(m_treeHit->GetBranch("posX")->GetAddress());
+  float pos_d;
+  DG_GET_DOTVALUE(&pos,&pos_d,sizeof(float));
+  std::cout << "pos_d=" << pos_d << "\n";
\end{lstlisting}
\end{lstfloat}

\begin{lstfloat}
\caption{Insertions into the source code of GATE for reverse-mode differentiation. Declare the input variable (beam energy) and output variables (hit positions). Additionally, the header \lstinline|derivgrind.h| must be included.}
\label{lst:insertion-reverse}
\vspace{0.2cm}
\begin{lstlisting}[language=c,frame=single,basicstyle=\ttfamily\scriptsize,linebackgroundcolor={\ifnum\value{lstnumber}>2\ifnum\value{lstnumber}<4\color{green!20}\fi\fi}]
   if (command == pEnergyCmd) {
     double energy = pEnergyCmd->GetNewDoubleValue(newValue);
+    DG_INPUTF(energy);
     pSourcePencilBeam->SetEnergy(energy);
   }
\end{lstlisting}
\begin{lstlisting}[language=c,frame=single,basicstyle=\ttfamily\scriptsize,linebackgroundcolor={\ifnum\value{lstnumber}>1\ifnum\value{lstnumber}<4\color{green!20}\fi\fi}]
   if (m_rootHitFlag) m_treeHit->Fill();
+  float pos = *(float*)(m_treeHit->GetBranch("posX")->GetAddress());
+  DG_OUTPUTF(pos)
\end{lstlisting}
\end{lstfloat}

Listing~\ref{lst:insertion-forward} shows the insertion of forward-mode client requests, which give 
access to the \emph{dot value} $\tfrac{\partial a}{\partial x}$ of any floating-point value $a$. 
In the first block, we set the dot value $\tfrac{\partial x}{\partial x}$ of the beam energy $x$ to~$1.0$. 
In the second block, we extract the particular output variable $f_i(x)$ of interest for AD, 
to retrieve and print its dot value $\tfrac{\partial f_i}{\partial x}$. 
Only the two modified source files of GATE need to be recompiled. Running GATE under Derivgrind 
reproduces the original output of GATE, interleaved with
additional output from Derivgrind and the sought derivatives.

Listing~\ref{lst:insertion-reverse} shows the insertion of recording-mode client requests to
mark $x$ as an input variable and $f_i(x)$ as an output variable for $i=1,2$.
Applied to the modified GATE program, Derivgrind records the real-arithmetic evaluation tree (\emph{tape}) and 
identifiers (\emph{indices}) for the input and output variables in the tree.
Reverse-mode AD is about tracking the \emph{bar value} $\tfrac{\partial f_i}{\partial a}$ of all floating-point
variables $a$ (here, for one output at a time). A simple tape evaluator program in the Derivgrind package can be used to allocate space for all bar values, set the bar value of the output variable $f_i$ to $1$, and evaluate the bar value $\tfrac{\partial f_i}{\partial x}$ of the input variable $x$ according to the tape. Reverse-mode AD finds the bar values of all input variables in one sweep and is therefore likely to provide a better run-time than numeric differentiation for optimization problems with many design parameters. 

\begin{table}
\caption{Numerical and automatic derivatives of $f_1$ and $f_2$ at $x_0 = \SI{230}{\mega\eV}$.}
\label{tab:derivatives}
\centering
\begin{tabular}{lcc}
\toprule
\multirow{2}[3]{*}{Differentiation method\qquad\qquad} & \multicolumn{2}{l}{approximation of $\frac{\partial f_i}{\partial x}$ in \si[per-mode=fraction,fraction-function=\frac]{\milli\meter\per\mega\eV}}  \\
\cmidrule(lr){2-3}
 &  \multicolumn{1}{c}{$i=1$} & \multicolumn{1}{c}{$i=2$}  \\ 
\midrule
Central difference quotient & & \\
\dots{} for $h=\SI{0.01}{\mega\eV}$   & $-0.0812531$ & $-0.130463$  \\
\dots{} for $h=\SI{0.005}{\mega\eV}$  & $-0.0811577$ & $-0.130272$  \\
\dots{} for $h=\SI{0.001}{\mega\eV}$  & $-0.0815392$ & $-0.131130$ \\
\midrule
Derivgrind, original \geant & & \\
\dots{} forward mode & $-0.0685116$ & $-0.113841$ \\
\dots{} reverse mode & $-1.72\cdot 10^8$ & $5.36\cdot 10^{13}$  \\
\midrule
Derivgrind, \lstinline|G4Log| $\rightsquigarrow$ \lstinline|log| & & \\
\dots{} forward mode & $-0.0813391$ & $-0.130524$ \\
\dots{} reverse mode & $-0.0813391$ & $-0.130524$ \\
\bottomrule
\end{tabular}
\end{table}

Table~\ref{tab:derivatives} lists the computed derivatives. The forward-mode automatic derivatives deviate from the difference quotients by about \SI{15}{\percent}, while the reverse-mode derivatives are completely off.
 Printing all the results of intermediate floating-point operations alongside their dot values and comparing with difference quotients, 
we were able to identify the statement at which they start to differ significantly. 
\geant defines an alternative math function \lstinline|G4Log| to numerically approximate the natural logarithm $\log z$ for $z\in\RR_{+}$, using an approximation algorithm adapted from the VDT math library~\cite{piparo_speeding_2014}. The algorithm starts with a range reduction step, multiplying the argument $z$ by $2^k$ for a suitable integer $k$ such that $\tfrac{1}{2} \leq z \cdot 2^k < 1$, by simply overwriting the exponent bits of $z$. Derivgrind does not recognize the real-arithmetic significance of this bit-trick.

Thus, we replaced \lstinline|G4Log| by a call to the standard C \lstinline+log+ function. GLIBC, and other implementations of the C standard library, also use bit-tricks to implement \lstinline+log+; however, Derivgrind recognizes and intercepts calls to C\,95 math functions, and uses the proper analytic derivatives. After this small code change in \geant, the automatic derivatives computed by Derivgrind's forward and reverse mode agree, and we indicated them by horizontal lines in Figure~\ref{fig:diffquot}. As these lines are surrounded from both sides by the markers indicating difference quotients, the automatic derivatives are either entirely correct, or at least their deviation from the true derivative is small compared to the variance of difference quotients. 

We have measured the run-times for a release-mode build of GATE/\geant with the \lstinline+time+ command on an exclusive node with two \SI{2.6}{\giga\hertz} Intel Xeon Gold~6126 processors at the University of Kaiserslautern-Landau's Elwetritsch cluster. The runtime goes up from around \SI{12}{\second} in native execution to \SI{13}{\min} in the forward mode, which is a factor of~65. Derivgrind's recording takes about \SI{24}{\min}, corresponding to a factor of~120, to record a tape of \SI{25}{\mega\byte} whose reverse evaluation takes about \SI{0.05}{\second}.

To summarize, the AD tool Derivgrind provides accurate derivatives for the parts of \geant analyzed in this study. Besides applying macros to input and output variables, the only change to the source code was to replace a alternative implementation of a math function by a call to the C math library. Therefore, it now becomes possible to include realistic Monte Carlo particle simulations into differentiable pipelines, instead of using surrogate models. Further research may compare these two approaches with respect to computational performance, convergence behaviour of the optimizer, and quality of the optimized designs.

\subsection{Optimization of derivatives using Polyhedral compiler}

Derivatives, mainly in the form of gradients and Hessians, 
are ubiquitous in machine learning and in frequentist and Bayesian inference. 

Traditionally, most AD systems have coded in high-level programs~\cite{pytorch, maclaurin_thesis},
and therefore have been unable to achieve a good performance on scalar code or memory-modifying loops. 
These systems have instead relied on extensive libraries of optimized kernels---\eg 
for linear algebra, convolutions, or probability distributions---combined with
associated adjoint rules, forcing practitioners to express their models in 
terms of these kernels to attain satisfactory performance.

New, low-level AD tools such as Enzyme~\cite{enzyme_NEURIPS2020} allow for differentiating
kernels implemented as naive loops. 
However, since the derivative code is generated programmatically during 
the application of reverse mode AD, the reverse passes are likely to access memory in 
patterns suboptimal with respect to both cache and SIMD performance. This opens a new 
opportunity for polyhedral loop-and-kernel compilers to provide the aggressive transforms 
needed for high performance.

In this section, we report on work in progress on the implementation of 
the LoopModels\footnote{Available at \url{https://github.com/JuliaSIMD/LoopModels}} automatic
loop optimization library based on LLVM Intermediate Representation (IR)~\cite{LLVM:CGO04},
which uses polyhedral dependency analysis methods to 
attain a performance competitive with vendor libraries on many challenging loop nests, 
with applications from linear algebra to the derivative code produced by AD.

\subsubsection{Techniques and applicability}

For many reasons, preserving performance optimisations during differentiation is a non-trivial task,
in particular in the context of reverse-mode automatic differentiation.
The footprint of read/write memory accesses of the generated derivative (adjoint) program usually differs 
drastically from that of the original program.
Implementation choices such as hand-tuned tiling strategies in the primal code are often 
not ideal for the derivative code.
While the overhead of suboptimal derivative programs is sometimes considered to be constant and non-significant,
it is often limiting in high-performance, demanding applications.

The LoopModels library will provide a compiler pass based on polyhedral modeling techniques, 
which will perform source code optimizations and autonomously certify the correctness of transformations.
This will allow the automatic optimization of adjoint programs, 
and hopefully alleviate the need for expert human intervention.
Ultimately, the library should allow machine generation of optimal 
code even for challenging cases like reverse-mode automatic 
differentiation of expressions containing highly-nested loops. 
By working at the LLVM IR level, LoopModels will naturally 
benefit from existing solutions in the LLVM compiler toolchain,
thus generating high-performance gradients of any language going to 
the LLVM intermediate representation, such as C/C++, Fortran, Julia, Rust, and others,
with a small overhead at compilation time.

LoopModels will rely on polyhedral methods for analyzing the loop programs.
The polyhedral model techniques for compiler optimization provide a powerful mathematical 
framework to represent nested loop computation and its data dependences using integral points in polyhedra.
This approach works by finding beneficial affine code transformations through a 
practical cost function that enables efficient fusion and tiling of arbitrarily 
nested loops in a synthesised adjoint program.
This allows simultaneous optimization for coarse-grained parallelism and locality. 

\subsubsection{Simple example}

Let us consider, as an example, a nontrivial loop transformation that 
LoopModels would be capable of discovering and applying.
The convolution operation is key to many 
applications and is one of the building blocks in machine learning (ML), and in particular deep learning (DL), pipelines.
When used in a convolutional neural network (CNN), 
the backpropagation stage in AD also requires the calculation of the gradient 
of the convolution with respect to its arguments during the reverse pass.

We examine the 1-D case, where given vectors 
$A \in \mathbb{R}^{N + I}$, $B \in \mathbb{R}^I$, and $C \in \mathbb{R}^{N}$, 
the convolution of $A$ and $B$ is defined for all $0 \le n \le N-1$ as: 
\begin{equation*}
    C_n = \sum_{i = 0\ldots I-1} A_{n + i} \cdot B_i\,.
\end{equation*}

Note that zero-based indexing is used here.
The pseudocode in Listing \ref{lst:alex-example-function-1} provides a 
basic implementation of the convolution operation. 

\begin{lstfloat}[H]
\caption{Original program}
\label{lst:alex-example-function-1}
\vspace{0.2cm}
\begin{lstlisting}[mathescape, language=python,frame=single,basicstyle=\ttfamily\scriptsize]
    function convolution(C, A, B)
        N, I = length(C), length(B)
        for n in 0:N-1 do
            for i in 0:I-1 do
                C[n] += A[n + i] * B[i]
            end
        end
    end
\end{lstlisting}
\end{lstfloat}
\begin{lstfloat}[H]
\caption{Adjoint program}
\label{lst:alex-example-function-2}
\begin{lstlisting}[mathescape, language=python,frame=single,basicstyle=\ttfamily\scriptsize]
    function adjoint($\overline{C}$, $\overline{A}$, B)
        N, I = length($\overline{C}$), length(B)
        for n in 0:N-1 do
            for i in 0:I-1 do
                $\overline{A}$[n + i] += $\overline{C}$[n] * B[i]
            end
        end
    end
\end{lstlisting}
\end{lstfloat}

One of the possible outputs of reverse mode AD applied to the program of Listing \ref{lst:alex-example-function-1}
is displayed in Listing \ref{lst:alex-example-function-2}. 
The function corresponds to the backpropagation algorithm that computes the adjoint $\overline{A}$, 
given the adjoint $\overline{C}$ and $B$; in this example, we ignore differentiation with respect to $B$.

While in this case the forward pass is easy to optimize via register tiling, 
this is not the case for the adjoint: the index into $\overline{A}$ in the innermost loop is dependent 
on both loop induction variables $i$ and $n$, making it impossible to hoist 
these memory loads and stores out of any loops. This forces us to 
re-load and re-store memory on every iteration, requiring several 
additional CPU instructions per multiplication.

In Listing \ref{lst:alex-example-function-3}, a possible optimized adjoint program
is presented. The transformation uses the observation that we can 
re-index the memory accesses to $\overline{A}$ so that it depends on one loop only.
By introducing new loop induction variables and adjusting the loop boundaries, 
we can rewrite the inner loop in a way that allows a register-efficient tiled access pattern.

\begin{lstfloat}[H]
\caption{Optimized adjoint program}
\label{lst:alex-example-function-3}
\begin{lstlisting}[mathescape, language=python,frame=single,basicstyle=\ttfamily\scriptsize]
    function adjoint_optimized($\overline{C}$, $\overline{A}$, B)
        N, I = length($\overline{C}$), length(B)
        for w in 0:N + I - 2 do
            for j in max(0, w - N):min(I - 1, w)
                $\overline{A}$[w] += $\overline{C}$[w - j] * B[j]
            end
        end
    end
\end{lstlisting}
\end{lstfloat}

\subsubsection{First experiments}

To justify the potential applicability of code transformations implemented in LoopModels, 
we report on experimental results produced within a proof-of-concept implementation.
For comparison, we consider \pytorch~\cite{pytorch}, and two ML libraries 
implemented in the Julia language~\cite{Julia-2017}:

\begin{itemize}
    \item \lstinline|Flux.jl|, a Julia general-purpose library for ML that uses high-level AD tools;
    \item \lstinline|SimpleChains.jl|, a Julia library that uses handwritten programs for 
    computing adjoints and applies code transformations in the spirit 
    of the approach proposed in LoopModels.
\end{itemize}

The performance of \lstinline|SimpleChains.jl| was compared to analogues on small-dimensional
datasets. 
Results are looking promising so far: 
for example, on MNIST dataset \cite{deng2012mnist} with a LeNet architechture
the full training pipeline in \lstinline|SimpleChains.jl|
took 1.5 seconds vs. the 50 seconds in \lstinline|Flux.jl| and 15 seconds in \pytorch, 
respectively\footnote{See also the talk by Chris Elrod at \url{https://youtu.be/rfBYA1gZa6E}, last visited on February 2023}.

\clearpage
\section{Progress in Muography Optimization}
\label{sec:tomopt}
 
In this section we describe the software package \tomopt (\emph{Differential Optimisation of Muon-Tomography Detectors}), which is the first concrete effort within the MODE Collaboration to research and develop differential optimisation techniques for detector design. Rather than immediately attempting to tackle LHC-scale instruments, we instead opt for the simplified, but nonetheless useful, domain of muon tomography, where both the detectors and inference chains are more easily managed. We described the details of \tomopt in a recent publication~\cite{Strong:2023oew}. \tomopt is a highly modular Python-based package that provides the full suite of tools and resources required for the investigation of the general problem of optimization of a scattering tomography detector. 

\subsection{Muon tomography}
Muons, elementary particles related to the electrons but about 200 times heavier, are produced by cosmic-ray interactions in the atmosphere. Their flux at sea level is of the order of \SI{100}{\hertz\per\metre\squared}, and their energy spectrum is very broad, peaking at a few \si{\gev} and extending up to the \si{\tev} scale. 
In the energy range \SIrange{1}{100}{\gev}, muons mostly loose energy by ionisation, at a rate of about \SI{200}{\mev} per meter of water. This makes them the most penetrating charged elementary particles. When traversing a material, muons undergo several elastic electromagnetic interactions with the nuclei of the traversed material (\emph{multiple scattering}),  As the strength of each collision depends on the charge of the nucleus, the deflection of a muon trajectory has a known dependence on the atomic number Z~\cite{Rutherford1911,LynchDahl1991} of the traversed material. This dependence can be inverted, to infer the atomic number of an unknown material by measuring the scattering angle of a batch of muons that scatter through it. The measurements are typically performed by means of two groups of layers of muon detectors, one above and one below the passive volume to be scanned. The muon trajectory above and that below are fitted using the hits generated in the layers by the muon passage, and the scattering angle between the two directions can be measured. \tomopt models this process in a differentiable pipeline where the detector parameters can be optimized by minimizing through AD-powered gradient descent a loss function that includes both the physics goal of the experiment and the cost of the detector configuration.

    \subsection{Package overview}
        \tomopt is built as a modular and user-inheritable Python package, backed by \pytorch~\cite{pytorch}. It is currently under development, with an open-source release planned soon, along with accompanying dedicated publications.

        The package implements all aspects of the simulation, detection, inference, and optimisation without external heavy dependencies. However, given the wide variety of possible applications, these aspects are presented as base classes, which are designed to be inherited by users and configured for their exact use-cases.

    \subsection{Usage}
        \tomopt is designed to iteratively adjust a detector system such that it becomes optimal; where optimality is quantitatively defined though the minimal value of a task-specific loss function that depends on the detector parameters.

        When setting up a problem, users define detector panels with an initial position and size, and also specify the dimensions of the passive volume to be imaged. Next, users must provide both a differentiable inference method, and a loss function. The former is used to provide predictions on properties of the passive volume, and the latter quantifies the error on these predictions. The exact nature of both of these methods will be dependent on the users' tasks, but \tomopt provides starting base classes for a range of problem categories. 
        
\begin{figure}[ht]
            \begin{center}
                \includegraphics[width=0.8\textwidth]{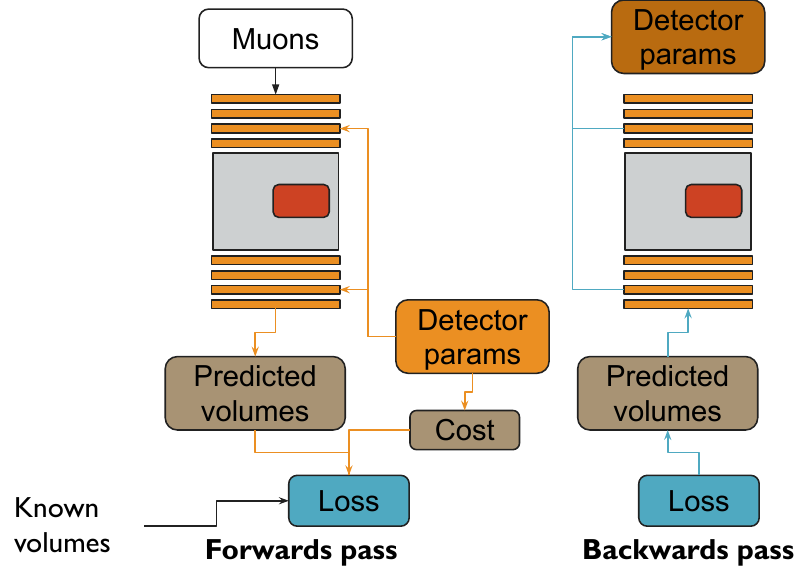}
                \caption{Forwards and backwards passes in \tomopt}
                \label{fig:tomopt:fwd_bwd}
            \end{center}
        \end{figure}

        In order to optimise the detector parameters (sizes and positions),
        typical layouts for the passive volume are sequentially loaded and inferred on. These may either be manually specified by the user, or generated by a suitable function. Inference is performed per passive volume layout using batches of many muons. The loss function may then be computed over batches of several passive volume layouts. The use of a differentiable inference method means that the analytic effect of each detector parameter may be computed via back-propagation of the loss gradient, and the parameters iteratively updated via gradient descent, as illustrated in \autoref{fig:tomopt:fwd_bwd}

    \subsection{Example}
        To better describe the usage, let us consider a concrete task: we need to scan containers in order to search for smuggled uranium, which might be hidden amongst scrap metal.
        
        \paragraph{Loss function}
            We can consider this a binary classification exercise, in which each container belongs to one of two classes: contains uranium or does not contain uranium. For such a task we can use binary cross-entropy as a loss function:
            \begin{equation}
                \mathcal{L}\!\left(y,\hat{y}\!\left(\theta\right)\right) = -y\ln\!\left(\hat{y}\!\left(\theta\right)\right)-\left(1-y\right)\ln\!\left(1-\hat{y}\!\left(\theta\right)\right),
            \end{equation}
            where $y\in\{0,1\}$ are the true class labels, and $\hat{y}\!\left(\theta\right)\in\left[0,1\right]$ are the predicted labels based on detector parameters $\theta$.

\begin{figure}[ht]
                \begin{center}
                    \includegraphics[width=0.8\textwidth]{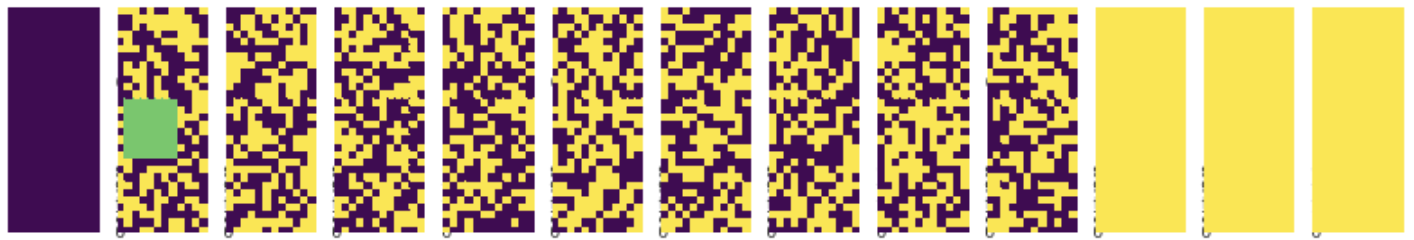}
                    \caption{Example: a passive volume in horizontal cross-sections starting from the bottom layer to the top. Yellow voxels indicate air, blue scrap metal, and green uranium.}
                    \label{fig:tomopt:passive}
                \end{center}
            \end{figure}

        \paragraph{Passive volumes}
            In order to simulate the passive volumes, we can generate examples by filling a metal container with a randomly varying amount of assorted metal, inter-spaced with air. The top of the container is also filled with air. With a specified probability, we can possibly then place a block of uranium of random shape inside the container, at a random location. An example volume is shown in \autoref{fig:tomopt:passive}.
    
        \paragraph{Detectors}
            We define detectors as panels placed parallel above and below the passive volume. When muons pass through the panels, hits will be recorded with a certain spatial resolution. These hits can later be used to infer the trajectory of the muon before and after traversing the passive volume. The detector panels each have five learnable parameters: position in $x,y,z$, and span in $x,y$.

            Since a physical detector-panel will either record a hit or not, depending on whether the muon passes through the panel, the hits would not be differentiable with respect to the $x,y$ parameters of the panel. To circumvent this problem, we instead model the detectors such that the resolution on the recorded hits varies with distance from the centre of the panel, and the span of panels. This means that the uncertainty in the true muon position varies as a function of the detector parameters.
            
            The realistic model for the detectors may still be used, when updates to the parameters are not required, e.g. when validating the detector configuration.

        \paragraph{Inference}
            When imaging a passive volume, many muons will be used. While each muon acts independently, it is convenient to group muons together and perform their propagation and inference in parallel for computational efficiency.
            
            In our case, full inference will be a two stage process: first we will use the muons to construct a rough 3D image of the volume by approximating the density of each voxel in the volume. This initial stage is rather task-agnostic; the next stage of inference takes this image and uses a task-specific algorithm to map the density predictions to final predictions, which in our case is a single number between zero and one representing the probability that there is uranium somewhere in the volume.

            The first stage of inference involves fitting trajectories to the incoming and outgoing hits, considering the uncertainties in each hit as a weight in the fit; thus each trajectory is differentiable with respect to the detector parameters. The changes in trajectory may be used to compute a prediction of the density of material in which the muon scattered. The position of this scattering may be predicted using, e.g. the ``point of closest approach" method~\cite{POCA}, which extrapolates the incoming and outgoing trajectories inside the passive volume and assumes that a single scattering occurred at the point closest to both trajectories (they are not guaranteed to intersect). With a sufficient number of muons, this method can be used to build up an estimate of the 3D distribution of the density in the volume, although it will be highly biased due to the assumption of a single point of scattering per muon.

            For the second stage of inference, we effectively need to convert a 3D tensor of floats to a single float. Here, one may choose to use \eg a three-dimensional (3-D) CNN to classify the volume, but classical approaches are also possible. In our case, we can expect that, although the predicted densities will not correspond to the true densities of the materials in the volume, their distribution should contain several peaks, according to the number of materials present: either two (air and scrap metal); or three (air, scrap metal, and uranium). Since air and scrap metal will always be present, we actually only need to quantify the presence of a high-density peak, which can be done by considering the difference between the mean of the $m$ highest-predicted densities, and the mean of the $n-m$ lowest predicted densities. If this difference is large, then it indicates the presence of uranium. After a suitable rescaling and sigmoid normalisation, we arrive at our required single-float prediction for the whole volume.

        \paragraph{Optimisation}
            An estimate of the value of the loss function at the current detector-parameter points can be computed by predicting and inferring many passive volumes. Through the use of automatic differentiation, the partial derivatives of the loss can be computed with respect to each parameter. Using the standard gradient update rule, the detectors can be improved by making one step of length $\gamma$ in the direction of steepest descent:
            \begin{equation}
                \theta_{t+1} = \theta_t-\gamma\nabla\mathcal{L}\!\left(y,\hat{y}\!\left(\theta_t\right)\right).
            \end{equation}

            This is the most basic optimisation loop: depending on the tasks and approaches used, however, it may be beneficial to augment the optimisation, or to run ML models as part of the pipeline. \tomopt enables such possibilities through the use of a stateful callback system, which allows classes to interject during the optimisation loop and have full read/write access to all aspects of the fit.

        \subsection{Status and prospects}
            As mentioned, \tomopt is still under development, but we intend to release it open-source soon, along with documentation. Two accompanying publications are planned: the first introduces the package from a more technical perspective, and demonstrates its application to an industrial example (ladle furnace), and has been recently released~\cite{Strong:2023oew}; the second focuses more on various possible approaches to inferring information about the passive volume from the scattering of muons.
            
\clearpage
\section{Progress in Calorimetry Optimization}
\label{sec:calopt}

Calorimetry is often the crucial part of a particle detector. By relying on destructive interaction of energetic particles with thick layers of matter, and the production of showers of secondaries, these devices are relatively simple in their functioning, yet the conversion of their output signals into physics measurements is made very complex by the stochasticity of the involved processes. Calorimeters have marked the history of particle physics in the past decades, and their continuous improvement has been a significant driver of new discoveries---it suffices to mention the detection of Higgs boson decays from measured photon pairs by the CMS and ATLAS Collaborations. 

Whereas the main task of both hadronic and electromagnetic calorimeters has been for a long time the one of detecting the collective energy yielded by all secondary particles produced in their interior, with relatively little emphasis on retaining or extracting precise position information about the energy depositions, these instruments withstood a transformation in recent times, when several physics-driven requirements (sticking with high-energy physics examples, it is unavoidable to mention here the separation of single photons from background-produced photon pairs in the case of the search for the Higgs boson, as well as the reconstruction of hadronic decays of boosted heavy objects in high-energy searches for new physics) have brought us to increase the transversal as well as the longitudinal segmentation of the active detection components. Consequently, the asymmetric nature of the development of particle showers naturally begs the question of what is the optimal arrangement and segmentation of calorimeter cells. Furthermore, new technologies nowadays allow to record the timestamp of energy depositions with an accuracy sufficient to lend itself as a fourth dimension to be studied and optimized. Together, these new capabilities also pose new questions on the possibility to actually exploit the difference in how different hadrons interact in dense media, with a view to extract particle identity information and further improve the particle-flow-based holistic reconstruction of complex hadronic showers typical of the big LHC experiments. In this section we consider a few use cases of relevance to the above program. 

\subsection {Optimization of the LHCb Calorimeter for the LHC phase 2 upgrade}

Optimization of a calorimeter refers to the development of a new or the modernization of an existing one.
In the case of an existing calorimeter, fine-tuning also involves certain constraints, due to the reuse of already existing components, which may instead be considered free parameters when studying a new development.
Examples of fine-tuning are the particular technology, geometry, and configuration of the calorimeter.
Typically, \emph{ab initio} ML approaches show that this fine-tuning can be avoided and a comparable reconstruction quality as in classical methods can be achieved~\cite{Boldyrev:2020ydy}.
When developing a model for the reconstruction of a real detector, both the classical and ML-based approaches require some preprocessing of the data to obtain geometry-agnostic inputs to the model.

As a first example of calorimeter optimization, we hereby consider the possible upgrade of the LHCb electromagnetic calorimeter (ECAL)~\cite{LHCb:2018roe}.
The current ECAL is based on Shashlik-type modules with transverse dimensions of 12$\times$12 cm$^2$~\cite{LHCb:2008vvz, LHCb:2014set}.
3312 such modules are arranged in a rectangular wall perpendicular to the LHC beam axis and are laterally segmented into 1, 4, or 9 cells.
In the upgraded calorimeter, some of the existing modules might be replaced by more granular modules that employ the SpaCal technology~\cite{Guz:2020mga}.
The response of a wall-like calorimeter in terms of reconstructed, analysis-level quantities can be represented as an image of an electromagnetic shower, where the value of each ”pixel” corresponds to the energy deposited in the corresponding cell of the calorimeter.
The lateral size of an electromagnetic shower can be determined by parametrization using the Molière radius specified by the technology.
In such an approach, it is possible to choose in advance the size of the considered area (\emph{window}) so that the specified fraction of all calorimeter clusters is contained within it, and the cell with the highest energy (\emph{seed}) is located around the center of the window.
In addition, by increasing the window size, one can apply algorithms that also estimate pile-up contributions, since in this case one can compare signal clusters and clusters from background contributions.
Most ML-based reconstruction algorithms in this approach are limited by the fixed dimensionality of the input array of energy deposits.
And if the window size is large enough, \eg 5$\times$5 cells, when scanning the entire calorimeter, we will find that many areas of the calorimeter will be inaccessible if we require that all cells in the window contain (homogeneous) information.
The calorimeter is therefore divided into three regions with three different granularities.

\begin{figure}
    \centering
    \includegraphics[width=0.9\linewidth]{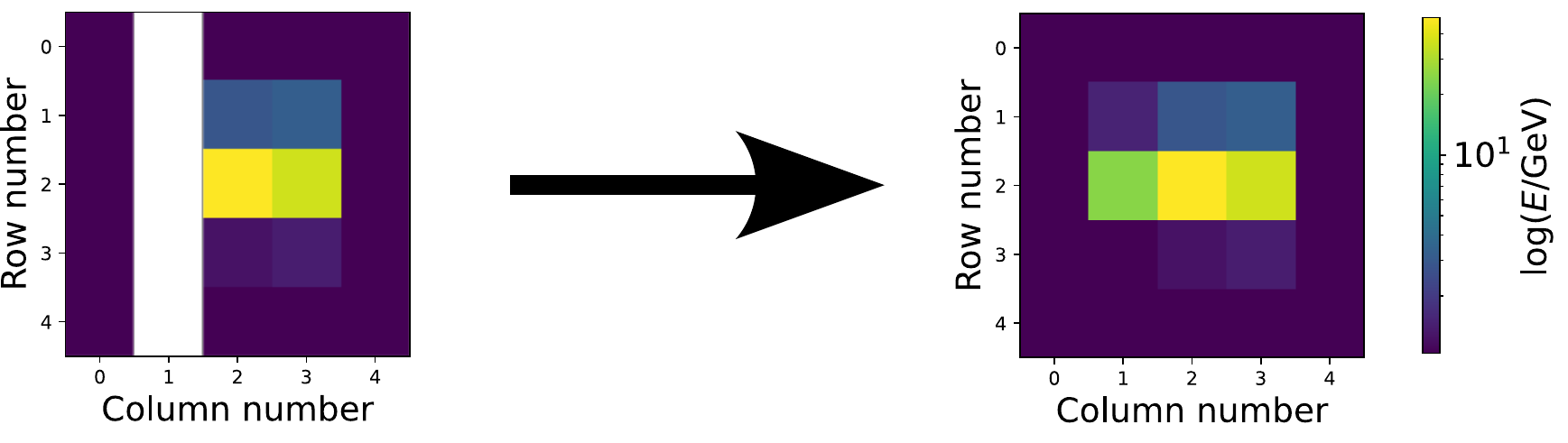}
    \caption{An example of a calorimetric cluster with missing column \#1 of cells (left) and a recovered cluster (right). The colour represents $\log(E/\mathrm{GeV})$ for each cell.}
    \label{fig:missing-cells}
\end{figure}

We consider different cases of deviations from the strict geometrical regularity of calorimeter cells arrangements.
In the first case, we will consider the boundary between calorimeter regions with different cell granularity.
In this case, it may be effective to divide large cells into smaller ones by interpolation.
In the second case, the irregularity may arise due to the technological necessity to rotate the modules slightly around one or two axes.
In this case, it is useful to use the coordinates of the cells as additional information at the input of the regressor.
In a third case, due to engineering reasons, it is possible to skip a row or column of cells:
one can then also interpolate the information in the missing cells (shown in Figure~\ref{fig:missing-cells}).
Finally, the irregularity at the borders of the calorimeter can be handled by padding.
By using such data pre-processing algorithms, we can significantly streamline the reconstruction model architecture.

Various interpolation techniques and DL models were compared to restore the missing rows or columns within the cell matrix.
The results obtained were averaged over the specific location of the missing row or column and are presented in Table~\ref{tab:ecal-interp}.
The best performing model, in terms of both peak signal-to-noise ratio (PSNR) and structural similarity index measure (SSIM), was found to be a fully connected neural network consisting of two linear layers activated by the rectified linear unit (ReLU).
We consider such pre-processing as part of a geometry-agnostic reconstruction model.
Such a model can be automatically trained on both simplified data sets designed for preliminary evaluation of calorimeter performance and data sets derived from detailed simulations.
The use of automatic training ensures consistency and uniformity of the reconstruction results.

\begin{table}
\caption{A comparison of classical interpolation methods and DL interpolation for the reconstruction of calorimetric clusters with missing information. Peak signal-to-noise ratio (PSNR) and structural similarity index measure (SSIM) metrics are used to measure the quality of cluster reconstruction. The results are averaged over the position of the missing row or column of cells. The best results are displayed in bold.}
    \label{tab:ecal-interp}
\centering
\begin{tabular}{cl ll l}
\toprule
\multicolumn{2}{c}{Model} & PSNR↑ & SSIM↑ \\ \midrule
\multirow{3}{*}{Interpolation} & Nearest-neighbor & 85.3 & 0.74 \\
 & Cubic & 91.2 & 0.75 \\
 & Linear & 92.8 & 0.79 \\ \midrule
Deep Learning & Fully-Connected & \textbf{96.9} & \textbf{0.94} \\
\bottomrule
\end{tabular}
\end{table}

\subsection {The challenge of beam-induced backgrounds in the electromagnetic calorimeter for a muon collider detector}
\label{sec:mucol}

The construction of a detector to study high-energy muon-muon collision poses significant challenges, many of which are entirely novel. One such case is provided by the very significant background due to in-flight decays of beam particles in the vicinity of the collision point. The detector is thus expected to be showered with a huge flux of low-energy photons and neutrons resulting from the interaction of decay products with structures around the collision area. This has been preliminarily taken into account at the machine-detector interface phase by introducing a tungsten nozzle \cite{Bartosik_2020}, which screens a big portion of the radiation, leaving almost exclusively the background coming from the area around the interaction point.
The chosen design is the one from Crilin \cite{Ceravolo_2022}: a dodecahedron, every edge of which is made of 5 layers of arrays of PbF2 cells - each equipped with silicon photomultipliers. The modular structure of this design lends itself quite naturally to geometrical optimization studies, and therefore is chosen as reference for our work. 

\begin{figure}
    \centering
    \includegraphics[width=0.9\linewidth]{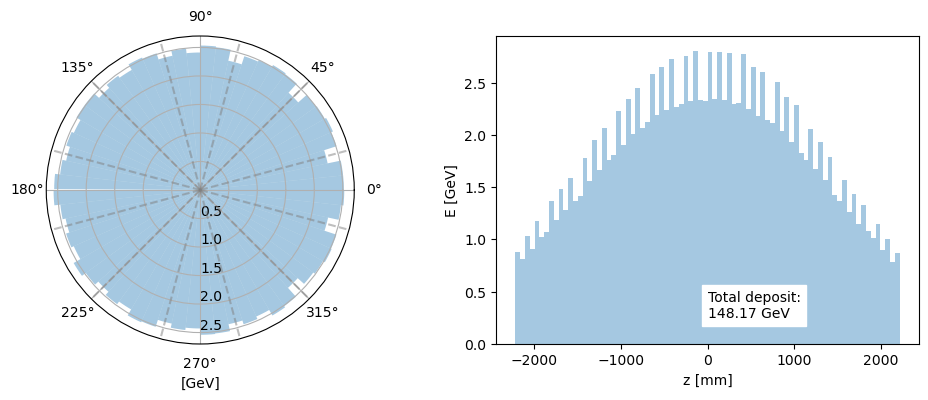}
    \caption{BIB deposition in PbF2 <ECal cells: Left: radial distribution with respect to beam axis (here defined as $z$), the dashed lines mark the calorimeter edges. Right: Distribution along beam axis. }
    \label{fig:BIB_ecal}
\end{figure}

This "Beam-Induced Background" (BIB) invites a revisitation of the construction paradigms of the detector components. In particular, the BIB is expected to significantly affect the detection of photon and electron-induced showers in the electromagnetic calorimeter. Figure \ref{fig:BIB_ecal} shows the deposition of a \geant-simulated \cite{GEANT} BIB event inside the calorimeter at a center-of-mass energy of 3TeV. The considerable amount of energy still left within the detector makes this a significant background that cannot be neglected. Furthermore, its asymmetric deposit distribution suggests that a uniform layout of the calorimeter cells would result in significant loss of performance with respect to a design that optimally adapted to the BIB flux geometry. 

While our studies of this particular application of calorimeter optimization are still in an initial phase, we mention it here for its special interest and for the potentially large impact that a full optimization may have. The work plan includes the following tasks: \par

\begin{itemize}
    \item Start with the choice of active material (PbF2
    is presently suggested) and geometry of the initial design for the central electromagnetic calorimeter;
    \item develop a continuous model of the BIB energy deposition as a function of depth in the calorimeter material and position with respect to beam axis and detector center. This must rely on a \geant simulation of the BIB particles as a function of energy and angle of incidence in the detector material;
    \item construct a function that returns the expected energy deposit in a given volume;
    \item create a model of the photon clustering and energy reconstruction;
    \item consider the three-dimensional size $\Delta x, \Delta y, \Delta z$ of calorimeter cells as parameters of a continuous model of the detector volume;
    \item generate real photon signals overlaid with BIB energy deposits using the parametrized model, reconstruct the signals, extract suitable metrics of utility;
    \item modify the size of calorimeter cells by following the gradient of the utility function, and iterate to convergence.
\end{itemize}

This procedure must, at the bare minimum, be complemented with the comparison of performance attained by a state-of-the-art photon reconstruction algorithm, given initial and final parameters. Higher accuracy can be sought for by producing maps of fast versus complete reconstruction performance.

\subsection {Optimizing irregular geometries}
An optimal calorimeter design, or future detectors in general, does not necessarily follow two dimensional grid structure, or structures that can be transformed trivially to fit into regular grids. Furthermore, for an optimisation of the detector from the material choices, the geometry, and down to the physics output, the reconstruction algorithms need to be capable of processing and utilising the correlations across the detector beyond regions of interest defined by seeds. On the other hand, the raw detector data is typically sparse, with only a small fraction of cells being active in each event. This is particularly true for hadronic showers in highly granular calorimeters or tracking devices.

The solution to these problems is two-fold: the data representation needs to be more generic, and the algorithms need to be capable of processing such data---conceptually, but also in terms of resource requirements. Currently, the most convenient way to represent the data is as a generic point cloud, with each point corresponding to a detector signal above threshold that can carry additional features such as the position, the shape, the deposited energy or even a full signal pulse shape. 
To process the point-cloud data, graph neural networks~\cite{ZHOU202057,velivckovic2023everything} are well suited, as they do not enforce a particular sorting of the points, yet provide information exchange across all points that can be used to infer the properties of the particles that the detector hits originated from. On the other hand, the choices are restricted by the resource constraints imposed by the hardware and the need to evaluate such models many times in a detector optimisation task. For these resource reasons, fully connected graphs, where connections scale with the number of points squared, are not feasible in the context of a whole detector optimisation. Therefore, graph neural network variants that are capable of learning the graph topology while keeping within resource constraints even for $\mathcal{O}(10^5)$ inputs are of particular interest \cite{wang2019dynamic,GravNet}. 

Not only the input dimensionality poses challenges, but also the fact that the large amount of detector cells originated from an unknown number of sparsely or densely distributed particles that entered the calorimeter. 
In absence of regular grids, well-defined outer edges of physics objects, and the possibility of defining meaningful bounding boxes in the physical space for particle reconstruction, classic object detection techniques from computer vision are not applicable. Instead, the object condensation formalism~\cite{Kieseler:2020wcq} is being employed for a growing set of reconstruction tasks based on graph neural networks~\cite{Qasim:2021hex, Qasim_2022, Bhattacharya_2023,dibello2022reconstructing}.

Based on these techniques, significant progress has been made in the past years, from first studies with $\mathcal{O}(10^4)$ inputs in a simplified environment~\cite{GravNet}, to the reconstruction of multiple particles in toy calorimeters with $\mathcal{O}(5\times10^4)$ inputs, as well as the CMS HGCAL~\cite{Bhattacharya_2023}, to point cloud sizes up to $\mathcal{O}(2\times10^5)$~\cite{Qasim_2022} within seconds, while maintaining promising physics performance, often outperforming classic approaches.

These algorithms can provide the basis for a calorimeter optimisation beyond grid-like structures, and open the possibility of a differentiable generic reconstruction algorithm for a full detector, allowing to jointly optimise different subsystems. 
The second ingredient for such an optimisation is finding differentiable surrogates for point cloud generation, which is a challenging task. Exploratory studies are currently ongoing and are reaching higher complexity at each iteration~\cite{kansal2022particle,buhmann2023epicgan,leigh2023pcjedi,buhmann2023caloclouds}. So far, these algorithms are not capable of simulating point clouds of the size that reconstruction algorithms can process, but the progress is promising: soon, studies on optimising more complex, not necessarily grid-structured detector designs with high granularity will become possible. Furthermore, even for grid-structured geometries, these reconstruction and generation algorithms can make very high granularity computationally feasible by exploiting the sparsity of the data.

\subsection{Optimization of the CMS High Granularity Calorimeter}

A case study is proposed for the optimization of the readout optical fibre plant of a high granularity calorimeter with over 6M channels. The High Granularity Calorimeter of CMS (HGCAL) is being designed for Phase II of the LHC and will cover the forward region of 1.5--3.0 in pseudo-rapidity~\cite{HGCAL-TDR}.
Given the dense pileup environment foreseen in the forward region, resulting from up to 200 simultaneous proton-proton collisions and the high number of channels measuring energy and time of particle showers, it is expected that a large event size, of the order of 4-6~MB is produced after each bunch crossing. The readout is expected to occur at a rate of 750 kHz.
A balanced throughput in the optical fibres as well as efficient aggregation of multiple fibres in bundles is required to maximize the usage of resources of the back-end electronics. Mechanical constraints are unavoidable in the routing of fibres, with break-points foreseen at the edge of each layer or the detector, where a re-arrangement (splicing) can occur.
In this contribution we have detailed these constraints and summarized the results obtained by a simple scan of the phase space. This simple approach allowed us to reduce the so-called dark fibre presence resulting in a significant decrease of the cost. The implementation of a ML algorithm poses however challenges due to the discrete nature of the problem. The example was presented as it could be a good use case for many other detectors with optical path continuum and discrete distribution of fibre patch panels.

\subsection {Studies of granular calorimetry for future collider experiments}

High-granularity in hadron calorimeters offers a significant increase in the performance of the information extraction procedures from particle interactions with dense media. At particle colliders, its benefits stem from a two-pronged revolution that took place at the beginning of this century. The first prong is constituted by boosted jet tagging, which was developed when it was recognized that the signal of hadronically-decaying heavy particles ($W$, $Z$, and $H$ bosons, top quarks, and other massive particles decaying to hadrons which may be hypothesized in new physics models) could be successfully extracted from backgrounds if sub-jets could be identified within wide jet cones. The second prong is constituted by the success of particle flow techniques, which were instrumental to \eg increase the energy resolution of hadronic jets at the CMS experiment above the non-state-of-the-art baseline performance of its hadron calorimeter. Both boosted jet tagging and particle flow reconstruction rely on accessing fine-grained information on the structure of hadron showers.

Other observations of the benefit of high granularity for future HEP endeavours include a recent demonstration~\cite{muregression_cnn} that fine-grained hadron calorimeters allow the measurement of the energy of multi-TeV muons from the pattern of radiative deposits (to a 20\% relative resolution that does not degrade with muon energy), offering itself as an obvious substitute to magnetic bending, which becomes impractical above a few TeV. If we want to preserve the discovery potential of energetic muons, this becomes an important aspect in the design of detectors at, \eg, a future circular collider for protons at energies above that of the LHC.

Jointly with, and independently from, the open hardware question of how far can granularity be pushed with existing or future available technologies, there remains an open question of how useful it can be, from an information extraction standpoint, to arbitrarily increase it. In principle, fine-granularity calorimeters offer more information than what we currently exploit them for. The nuclear interactions that a proton, a pion, or a kaon withstand when they traverse dense matter are different in cross section as well as in outcome, and this difference corresponds to information which until today we have never even attempted to extract. DL algorithms may today allow it, if only in probabilistic terms that are still going to be strongly useful for particle flow reconstruction and AI-based pattern recognition. The question to be investigated is whether this information extraction is feasible.

\begin{figure}[ht!]
\includegraphics[width=0.9\linewidth]{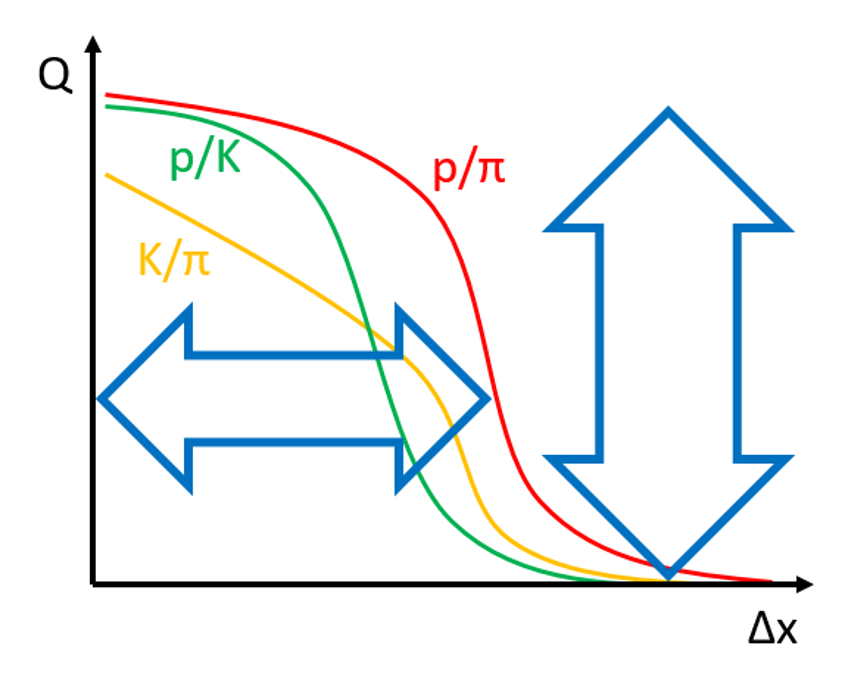}
\caption{Schematic view of possible discrimination power $Q$ for pairs of hadrons as a function of the calorimeter cell size $\Delta x$. Of interest are both the value of $Q$ attainable with arbitrarily high granularity (vertical arrow), and the value of $\Delta x$ above which no significant discrimination can be extracted from data (horizontal arrow).}
\label{f:caloID}
\end{figure}

One of the first issues to address is to provide a specific quantification of the following question: What are the ultimate particle identification capabilities of an arbitrarily granular hadron calorimeter? In more quantitative terms, this question can be turned into the quantification of two sets of numbers. The first set is composed by the highest achievable Bayes factors $Q$ of hypothesis testing for discriminating long-lived hadrons---\eg, protons from charged kaons, protons from pions, pions from kaons---assuming no lower limit on the size $\Delta x$ of individually readout cells. The second set of numbers then determines the feasibility of such an instrument, through the physical size of cells above which any meaningful information (practically useful $Q$ values) is lost (see Fig.~\ref{f:caloID}). A meaningful determination of these quantities requires the deployment of state-of-the-art DL models and very careful studies, and possibly also an extrapolation to future capabilities of those models, as what is of specific interest is precisely the technical limit of such discrimination performance. Of course, the answers will depend on the details of the chosen technology for signal detection and active material of the calorimeter, which adds an additional dimension and further interest to this study. 

It should be noted that despite the intrinsic interest of determining the quantities discussed above, they are only intermediate proxies to any final goal of a measuring instrument, and thus only a preliminary step (albeit a quite informative one at that) in the optimization task of a calorimeter. They would be crucial inputs to define the range of parameter values for the design of an instrument in any specific application.

Combined with the above research questions, one should also consider how much additional information gain is available by exploiting timing information in this high-granularity setup, and what are the potential performances on strange-quark tagging in a future collider by combining the time-of-flight discrimination with the spatial one.

If we find out that there is valuable information in the primary particle identity, which can be mined in the fine-structure of the development of hadronic showers, and that this results in a corresponding significant gain for future instruments, the maximum granularity that is capable of retaining it becomes the gauge with which to measure whether our technology can be exploited for the task. Timing information can then be studied as the necessary complement, given recent developments in ultra-high time resolution.

The studies needed to satisfactorily address the above questions require, in a first phase, the deployment of large DL models trained and tested on large simulated datasets; and the prototyping and test-beam operation of a small-scale demonstrator if the simulations demonstrate potential for hadron ID separation for cell sizes that are---or that may potentially become in the near future---technologically feasible.

\subsubsection {Hybridization of tracking and calorimetry}

A long-standing paradigm in detector design for HEP can be summarized by the motto “track first, destroy later”. With no exception, particle tracking has been reliant on low-material-density to avoid as much as possible the degrading effect of nuclear interactions; and conversely, calorimetry has exploited dense materials for efficient energy conversion in limited volumes. However, the advent of DL questions the validity of that paradigm, as today’s neural networks can make sense of the complex patterns resulting from nuclear interactions. It appears therefore highly desirable to investigate the possibility to trade off some of the undeniable benefits of light-weight tracking (in terms of resolution and low background) for a better reconstruction of the identity and development of hadronic jets. The abovementioned hypothesized possibility to discriminate the identity of different particles based on their behavior in traversing matter invites a study of what are the performance gains and losses of a combination of a state-of-the-art tracker followed by a fine grained calorimeter, when the density of the former and the latter do not abruptly change at their interface, but rather vary with continuity from the first to the second. Beyond the possibility of particle identification, a hybridization of tracking and calorimeter brings in a natural impedence matching with the state-of-the-art of particle flow reconstruction, in the sense that it potentially provides the algorithms with a larger and more coherent amount of information about the behavior of individual particles and their interaction history within showers.

The study of the above subject is again reliant on the deployment of highly specialized DL models, and in fact it requires to extrapolate to the future capabilities of these algorithms to the time when such a detector could become operational. It could be articulated as follows:
(1)	Study ultimate performances, on specific high-level benchmarks (e.g., precision of the extraction of a H$\to$bb signal in specific FCC or Muon-collider setups) of a idealized state-of-the-art tracker plus calorimeter (e.g. starting with an existing design, such as the CMS central detector), with a developed DL reconstruction.
(2)	Consider increasingly hybrid scenarios when the outermost layers of the tracker are progressively embedded in the calorimeter, gauging the performance on low-level primitives (single-particle momentum resolution, fake rates) and high-level objectives.
Eventually, such a study should inform the one described above in (1), to converge on a design of a future instrument capable of optimally exploiting the enhanced information extraction potential.
(3)	End-to-end optimization: the studies included in the above project will inform a full modeling by differentiable programming of the whole chain of procedures, from data collection to inference extraction, which allows to directly connect the final utility function of an experiment with its design layout and technology choices, such that the navigation of the pipeline by stochastic gradient descent may allow a full realignment of design goals and implementation details. 

\clearpage
\section{Progress in Accelerator Applications}
\label{sec:accel}

Particle accelerators are a critical part of enabling discoveries in high-energy physics. The goal of accelerator science is to advance our understanding of fundamental beam physics and particle accelerators while developing novel methods and tools to aid in the operation of current beam facilities and the development of future ones.  

High energy physics applications often require beam distributions in a 6-D position-momentum phase space ($x,p_x,y,p_y,z,p_z$) that are tailor-made for individual applications. 
For example, beams must be compressed longitudinally and flattened transversely to improve collider luminosity~\cite{phinney_ilc_2007}, controlled transversely to mitigate beam losses in high-intensity accelerators~\cite{ball_pip-ii_2017}, or shaped longitudinally to mitigate emittance growth due to coherent synchrotron radiation~\cite{mitchell_longitudinal_2013} and improve the performance of novel acceleration techniques~\cite{colby_roadmap_2016, seryi_concept_2009, tzoufras_beam_2008, roussel_single_2020}.
Manipulating beam distributions at a fine level represents a paradigm shift from traditional beam dynamics control goals, which only seek to control high-level beam properties.
This level of control requires diagnostic techniques that provide a similarly detailed reconstruction of the beam distribution in the 6-D phase space, far beyond traditional~\cite{wiedemann_particle_2007} or more recent~\cite{prat_four-dimensional_2014} approaches that infer only scalar properties of the beam.

Enabling practical detailed reconstructions of 6-D phase space distributions requires novel techniques that reduce diagnostic and numerical complexities associated with current methods.
For example, pinholes~\cite{gordon_four-dimensional_2022}, slits (combined with longitudinal phase space manipulations)~\cite{cathey_first_2018}, mesh grids~\cite{marx_single-shot_2018} or laser wires~\cite{wong_4d_2022} have been used to provide high-dimensional ($>$ 2-D) information about the beam distribution.
However, these techniques require specialized diagnostic elements and/or a large number of measurements to produce high-resolution reconstructions of the beam distribution. 
On the other hand, tomographic manipulations of the beam distribution in phase space, combined with commonly available detailed measurements of 2-D transverse beam distributions at diagnostic screens, have also been used to reconstruct high dimensional phase space distributions~\cite{sander_beam_1980, wang_four-dimensional_2019, wolski_transverse_2020}.
Unfortunately, these tomographic reconstruction techniques incur in significant computational costs when trying to infer high dimensional distributions from 2-D projections.
Finally, while ML techniques have been implemented to reconstruct phase space distributions from experimental data~\cite{scheinker_adaptive_2021, wolski_transverse_2022}, they demand significant initial investment to be effective, including the generation of large training data sets from simulation or experiment and the training of ML models.
These limitations of detailed, high-dimensional measurement techniques hinder their practical usage, restricting our comprehension and control of beam dynamics within accelerators.

\begin{figure*}[ht]
    \includegraphics[width=\linewidth]{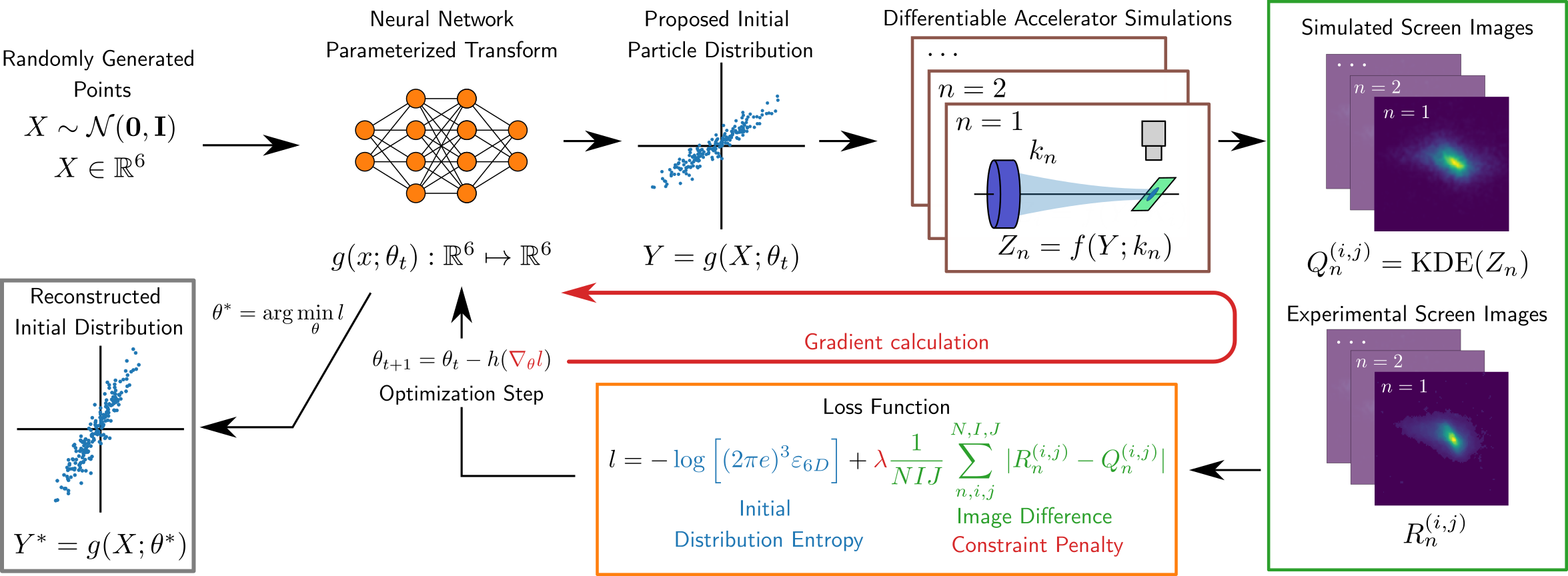}
    \caption{Description of the approach for reconstructing phase space beam distributions using differentiable accelerator physics simulations.
    First, randomly generated points drawn from a multivariate normal distribution $\mathcal{N}(\mathbf{0},\mathbf{I})$ are transformed via a neural network, parametrized by $\theta_t$, into a proposed initial distribution. This distribution is then transported through a differentiable accelerator simulation of the tomographic beamline. The quadrupole is scanned to produce a series of images on the screen, both in simulation and on the operating accelerator. The images produced both from the simulation $Q^{(i,j)}_n$ and the accelerator $R^{(i,j)}_n$ are then compared with a custom loss function, which attempts to maximize the entropy of the proposal distribution, constrained on accurately reproducing experimental measurements. Neural network parameters $\theta_{t} \to \theta_{t+1}$ are then iteratively tuned via gradient descent in order to minimize the loss function. Reproduced from Ref.~\cite{roussel_phase_2023}.}
   \label{fig:cartoon}
\end{figure*}

We have developed a novel method to provide detailed reconstructions of the beam phase space using simple and widely-available accelerator elements and diagnostics. 
To achieve this, we take advantage of recent developments in ML to introduce two new concepts (shown in Fig.~\ref{fig:cartoon}): a method for parametrizing arbitrary beam distributions in 6-D phase space, and a differentiable particle tracking simulation that allows us to learn the beam distribution from arbitrary downstream accelerator measurements.
This method extracts detailed 4-D phase space distributions from measurements in simulation and experiment, using a simple diagnostic beamline, containing a single quadrupole, drift and diagnostic screen to image the transverse ($x,y$) beam distribution.  

\begin{figure}[t]
    \includegraphics[width=\linewidth]{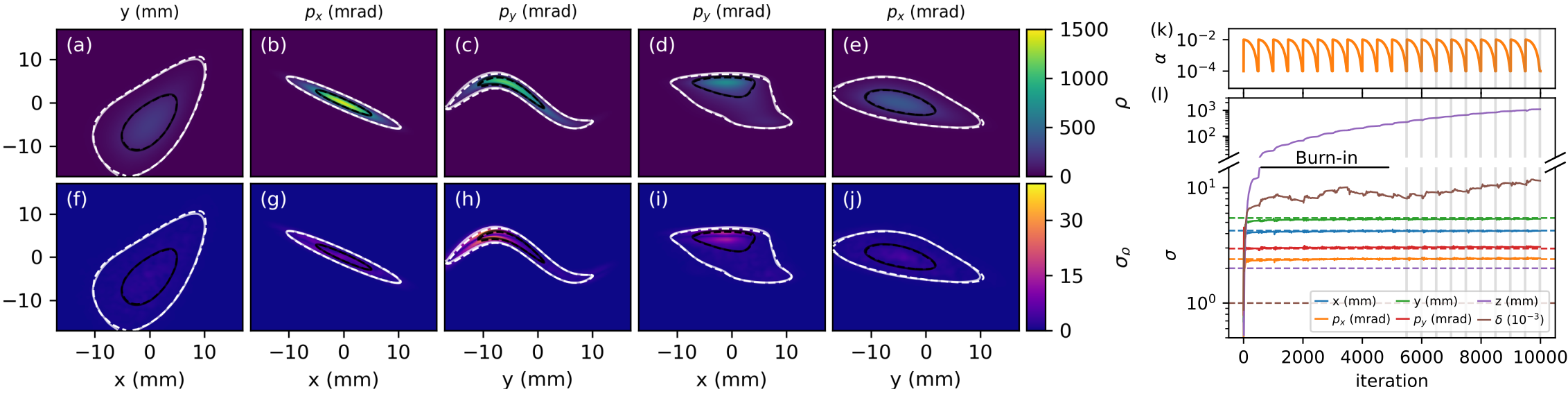}
    \begin{centering}
    \vspace*{-2mm} 
    \caption{Left: Comparisons between the synthetic and reconstructed beam probability distributions using our method. (a-e) Plots of the mean predicted phase space density projections in  4-D transverse phase space. Contours that denote the 50\textsuperscript{th} (black) and 95\textsuperscript{th} (white) percentiles of the synthetic ground truth (dashed) and reconstructed (solid) distributions. (f-j) Plots of the predicted phase space density uncertainty.
    Right: Evolution of the proposal distribution during training on synthetic data. (k) Learning rate schedule for snapshot ensembling. 
    (l) Second-order moments of beam reconstruction during training for each phase space coordinate. Dashed lines denote ground truth values. Vertical lines denote snapshot locations after burn-in period. Reproduced from~\cite{roussel_phase_2023}.}
    \label{fig:synthetic_results}
    \end{centering}
\end{figure}

We demonstrated our algorithm using a synthetic example, where we attempt to determine the distribution of a 10-MeV beam given a predefined structure in 6-D phase space. The propagation of a synthetic beam distribution through a simple diagnostic beamline containing a 10cm long quadrupole followed by a 1.0m drift is simulated using a custom implementation of the beam dynamics code Bmad~\cite{gonzalez_wepa065}.
To illustrate the capabilities of our technique, the synthetic beam contains multiple higher-order moments between each phase space coordinates. 
To simulate an experimental measurement, we simulate particles traveling through the diagnostic beamline while the quadrupole strength $k$ is scanned over $N$ points.
The final transverse distribution of the beam is measured at each quadrupole strength using a simulated $I\times J = 200 \times 200$ pixel screen, with a pixel resolution of 300\textmu m.
The set of images, where the intensity of pixel ($i,j$) on the nth image is represented by $R^{(i,j)}_n$, is then collected with the corresponding quadrupole strengths to create the data set, which is then split into training and testing subsets by selecting every other sample as a test sample, resulting in 10 samples for each data subset.

The reconstruction algorithm begins with generating arbitrary initial beam distributions (referred to here as proposal distributions) through a neural network transformation.
A neural network, consisting of only two fully-connected layers of 20 neurons each, is used to transform samples drawn from a 6-D normal distribution centered at the origin to macro-particle coordinates in real 6-D phase space (where positional coordinates are given in meters and momentum coordinates are in radians for transverse momenta).
As a result, the coordinates of particles in the proposal distribution are fully parameterized by the neural network parameter set $\theta_t$.

Fitting neural network parameters to experimental measurements is done by minimizing a fully differentiable loss function to determine the most likely initial beam distribution, subject to the constraint that it reproduces experimental measurements; this is similar to the MENT algorithm~\cite{Hock_A_2013}.
The likelihood of an initial beam distribution in phase space is maximized by in turn maximizing the distribution entropy, which is proportional to the logarithm of the 6-D beam emittance $\varepsilon_{6D}$~\cite{lawson1973emittance}.
Thus, we specify a loss function that minimizes the negative entropy of the proposal beam distribution, penalized by the degree to which the proposal distribution reproduces measurements of the transverse beam distribution at the screen location.
To evaluate the penalty for a given proposal distribution, we track the proposal distribution through a batch of accelerator simulations that mimic experimental conditions to generate a set of simulated images $Q^{(i,j)}_n$, which we then compare with experimental measurements. 
The total loss function is given by:
\begin{equation}
l= -\log \Big[(2\pi e)^3\varepsilon_{6D}\Big] + \lambda\frac{1}{NIJ}\sum_{n,i,j}^{N,I,J}|R^{(i,j)}_n - Q^{(i,j)}_n|\,,
\label{eqn:loss_fn}
\end{equation}
where $\lambda$ scales the distribution loss penalty function relative to the entropy term and is chosen empirically based on the resolution of the images. 

Results from our reconstruction of the initial beam phase space using synthetic images are shown in Fig.~\ref{fig:synthetic_results}.
We characterize the uncertainty of our reconstruction using snapshot ensembling~\cite{huang2017snapshot}.
During model training, we cycle the learning rate of gradient descent in a periodic fashion: this encourages the optimizer to explore multiple possible solutions (if they exist).
After several of these cycles (known as a ``burn-in" period), we save model parameters at each minimum of the learning rate cycle, as shown in Fig.~\ref{fig:synthetic_results}(a).
We then weight predictions from each model equally, using them to predict a mean initial beam density distribution Fig.~\ref{fig:synthetic_results}(a-e) with associated confidence intervals Fig.~\ref{fig:synthetic_results}(f-j).
Performing this analysis by tracking $10^5$ particles for each image took less than 30 seconds per ensemble sample using a professional grade GPU ($<60$ ms per iteration, 500 steps per ensemble sample).

We see excellent agreement between the average reconstructed and synthetic projections in both transverse correlated and uncorrelated phase spaces.
Furthermore, the prediction uncertainty from ensembling is on the order of a few percent relative to the predicted mean, providing confidence that the overall solution found during optimization is unique.
Finally, reconstructions of the beam distribution from image data predicts transverse phase space emittances that are closer to ground truth values than those predicted from traditional measurement techniques, i.e.\ second-order moment measurements of the transverse beam distribution.
This results from non-linearities and cross-correlations present in the 4-D transverse phase space distribution.

It is instructive to examine the evolution of the proposal distribution during model training.
In Fig.~\ref{fig:synthetic_results}(l) we examine second-order scalar metrics of the proposal distribution after each training iteration for each phase space coordinate.
The entropy term in Eq.~\ref{eqn:loss_fn} causes the distribution to expand in 6-D phase space until constrained by experimental evidence.
Phase space components that have the strongest impact on beam transport through the beamline as a function of quadrupole strength converge quickly to the true values, whereas the ones that have little-to-no impact (e.g. the longitudinal distribution characteristics) continue to grow.
In other cases, there is weak coupling between the experimental measurements and beam properties; for example, chromatic focusing effects due to the energy spread $\sigma_\delta$ of the beam weakly affect the measured images.
Here, the reconstruction can only provide an upper-bound estimate of the energy spread, since small changes in transverse beam propagation due to chromatic aberrations are overshadowed by statistically dominated particle motion.
Convergence of the proposal distribution thus provides a useful indicator of which phase space components can be reliably reconstructed from arbitrary sets of measurements.

The development of tools and techniques that leverage differentiable accelerator physics modeling has the potential to revolutionize key aspects of how experimental data is interpreted in the field of accelerator physics. 
Differentiable beam dynamics modeling directly addresses fundamental limitations facing traditional analysis approaches in accelerator physics. By enabling the combination of physical laws with the flexibility of ML-based function representations and detailed experimental measurements of accelerator beams, it can provide an unparalleled understanding of detailed beam dynamics inside accelerators.

\clearpage
\section{Progress in the Optimization of Experiments for Astrophysics Research}
\label{sec:astro}

\subsection{Dark Matter direct detection experiments}

Dark Matter (DM) particles pervading our Galactic halo could be directly detected by measuring their scattering in a suitable detector. The rare and small expected signal requires ultra-low background conditions and low energy detection thresholds. After summarizing the features of this possible DM signal and briefly describing the experimental efforts to detect it, we will outline the application to DM direct searches of ML techniques: these allow the discrimination of the expected signal from radioactive backgrounds or other noise events, being typically more effective than conventional filtering protocols. This capability is very important, taking into account the increasing demands to lower backgrounds and thresholds in future experiments.

\subsubsection{Dark Matter signals}
The presence of DM is required to explain an important fraction of the energy-mass budget of the Universe following different cosmological and astrophysical observations, although its nature is unknown~\cite{RevModPhys.90.045002}. A plethora of possible DM candidates have been proposed, being non-zero-mass, stable particles having a very low interaction probability with baryonic matter. Among them, thermal Weakly Interacting Massive Particles (WIMPs) are supposed to have been produced in the early Universe via a freeze-out mechanism when Standard Model (SM) and DM particles were in thermal equilibrium, producing a constant relic density, reproduced for a wide range of masses from 1~eV/c$^2$ to 120~TeV/c$^2$. 

Different complementary strategies are being attempted for WIMP detection~\cite{snowmass}. DM candidates could be produced at colliders and indirectly detected by identifying an excess of SM particles like gamma-rays, neutrinos, positrons or antiprotons produced by the annihilation of DM particles. In the direct detection of DM in the Galactic halo, the goal is to register the elastic scattering of WIMPs off target nuclei or electrons in a detector~\cite{Billard_2022}. Taking into account the expected signal from this interaction, the direct detection of DM is really challenging: the interaction has an extremely low probability, and large exposures and low background conditions (operating deep underground to suppress the effect of cosmic rays) are mandatory; the signal is concentrated at very low energies (below tens of keV), which requires the use of low energy threshold detectors; and the signal has a continuum energy spectrum, which would appear entangled with background, therefore distinctive signatures would be helpful to assign a DM origin to a possible observation.

Direct detection experiments can be focused on different physics cases~\cite{Billard_2022}; many of them just look for an excess of events over the known backgrounds, considering different ranges of candidate masses, Nuclear or Electronic Recoils (NR/ER) or different types of interactions between the dark matter particles and the nuclei (Spin-Independent, SI or Spin-Dependent, SD). Other experiments search specifically for distinctive DM signatures, like the annual modulation in the interaction rate or the directionality. Several physics cases are described in the following.
\begin{itemize}
\item There are particular requirements to probe DM candidates with masses at sub-GeV/c$^2$ scale: lighter targets must be used to keep kinematic matching between WIMPs and nuclei, lower threshold are necessary to detect smaller signals, and new search channels (absorption or scattering off by electrons, ER) are being considered, as light WIMPs cannot transfer sufficient momentum to generate detectable NR. Following the proposed Migdal effect\footnote{Atomic physics effect that leads to the emission of a bound-state electron from atomic or molecular systems when the atomic nucleus is suddenly perturbed. It has been observed for radioactive decays; there is no evidence for NR yet, although attempts are in progress.}, the DM-nucleus interaction could lead to excitation or ionization of the recoiling atom, being for low mass DM this additional signal above threshold (unlike the NR alone) and then enhancing sensitivity~\cite{ibe}; for this reason, this effect is already being considered by many collaborations to release results exploring sub-GeV masses. 
\item The movement of the Earth around the Sun makes the relative velocity between detectors and DM particles in the Galactic halo oscillate in time, which produces a modulation in the expected DM interaction rate with defined features like a one-year period; this signature would allow identifying a possible DM signal~\cite{PhysRevD.37.3388}. The DAMA/LIBRA experiment at the Laboratori Nazionali del Gran Sasso in Italy is observing for more than 20 years an annual modulation in the measured rate compatible with DM~\cite{dama2021}; this modulation signal has not been confirmed nor refuted at high confidence level by other experiments.
\item The average direction of DM particles through the solar system comes from the constellation of Cygnus, as the Sun is moving around the Galactic center; the measured track direction of NR could be therefore used to distinguish a DM signal from background events (expected to be uniformly distributed) and to prove the Galactic origin of a possible signal~\cite{PhysRevD.37.1353}. The main difficulty is to reconstruct the very short tracks ($\sim$1~mm in gas, $\sim$0.1~$\mu$m in solids) expected for keV scale NRs~\cite{Battat:2016pap}.
\end{itemize}

\subsubsection{Detection}
Many different and complementary technologies are being applied or under consideration in experiments attempting the direct detection of DM, \eg solid-state cryogenic detectors, time projection chambers based on noble liquids, scintillating crystals, and purely ionization detectors using semiconductors or gaseous targets; detectors measure the heat, light or charge produced or a combination of two of them in hybrid detectors. A discussion of advantages and disadvantages for each technique and relevant results obtained in the field can be found at~\cite{Billard_2022}. The properties of the DM candidates are constrained under different scenarios for the interaction. 

For high mass DM, experiments using large liquid noble detectors (Xe and Ar) provide now the best limits on cross-sections for SI DM-nucleus interaction and will explore regions of cross-sections where solar and atmospheric neutrinos become an irreducible background with projects using even larger detectors starting at the end of the decade~\cite{snowmasssnufog}; for SD DM-proton interaction, bubble chambers provide the best limits.

For low mass DM, the best sensitivity comes from a combination of experiments based on different detection techniques: solid-state cryogenic detectors (using scintillating bolometers or small mass Ge and Si semiconductor crystals), purely ionization detectors (Ge diodes, CCDs or gaseous detectors) and liquid noble detectors when operated in low threshold mode; candidates with increasingly lower masses could be investigated thanks to the development of novel technologies to further reduce the energy threshold~\cite{snowmasslth}. 

It is also worth noting that important results from NaI(Tl) experiments~\cite{PhysRevD.103.102005,PhysRevD.106.052005} to solve the long-standing conundrum of the DAMA/LIBRA annual modulation result have been presented and that studies for a DM detector with directional sensitivity are underway to prove the Galactic origin of a possible signal~\cite{cygnus,snowmassdir}.

\subsubsection{Machine-learning techniques}
In the context of DM direct detection experiments, ML techniques are being applied mainly to improve the discrimination of the expected signal from radioactive background or other type of noise events. A few examples of application of these techniques in this context are highlighted here.
\begin{itemize}
\item The sensitivity for low-mass DM searches of detectors of the EDELWEISS experiment (made of germanium cryogenic bolometers and operated in the Modane Underground Laboratory in France) has been studied in Ref.~\cite{PhysRevD.97.022003}. Using a data-driven background model, frequentist and multivariate analysis approaches (profile likelihood and boosted decision tree) are used to compute exclusion limits on cross-sections.
\item The LUX experiment used a dual-phase Xe TPC in the SURF laboratory in the US; backgrounds from the wire grid electrodes near the top and bottom of the active target were found to be particularly pernicious, limiting the sensitivity to low-mass DM. A ML technique based on ionization pulse shapes to specifically identify and reject these background events was developed and sensitivity improved~\cite{PhysRevD.104.012011}. Moreover, results  combining ML with the profile likelihood fit procedure using LUX data have been presented in Ref.~\cite{PhysRevD.106.072009} as a fast and flexible analysis of DM data; it is considered that this technique can be exploited by future DM experiments to make use of additional information and reduce computational resources needed for signal searches and simulations.
\item A new type of analysis for the DRIFT-IId directional DM detector (operated in the Boulby laboratory in UK) using a Random Forest classifier has been presented in Ref.~\cite{Battat_2021}. Events are labelled as signal or background based on a series of selection parameters, rather than solely applying hard thresholds, allowing an increased efficiency at lower energies and a projected sensitivity enhancement of even one order of magnitude for some WIMP masses.
\item For experiments using NaI(Tl) scintillators to search for a possible annual modulation in the interaction rate, raw data below $\sim$10 keV are fully dominated by non-scintillation events mainly related to the photomultipliers (PMTs) coupled to crystals; multiparametric thresholds on parameters deduced from the pulse shapes of the PMT signals (like first momentum or ratios of different fractions of the areas) are mandatory for careful event selection. But an unavoidable leakage of noise in the lowest energy bins can happen. Boosted Decision Trees (BDT) have been developed to improve the rejection of PMT-related noise with a multivariate analysis combining several weak discriminating variables into a single powerful discriminator. Decision trees have a binary structure with two classes, in this case associated to signal and noise, respectively. This approach has been followed by the COSINE-100~\cite{2021102581}, SABRE~\cite{sabre1,sabre2} and ANAIS-112~\cite{Coarasa_2021} experiments, obtaining lower energy threshold or background levels just above the threshold. In particular, in the case of ANAIS-112 the use of ML techniques has allowed to improve the sensitivity to the DAMA/LIBRA signal, as reported in Ref.~\cite{Coarasa_2022}; training populations independent of background data have been obtained from dedicated onsite neutron calibrations (producing genuine NR) for signal-like events and from data taken from a blank module (identical to the ANAIS-112 detectors but without a scintillating NaI(Tl) crystal) for noise-like events. The optimal thresholds on the BDT parameter (being -1 for noise and +1 for signal) have been defined for each energy bin and for each detector independently and the efficiencies for the selection of bulk scintillation events in the region of interest carefully estimated. Thanks to the improvement in background rejection in that region (a background level reduction of around 20\% has been achieved between 1 and 2 keV) and the increase in detection efficiency with this analysis with respect to the previous filtering, the ANAIS-112 sensitivity to test the DAMA/LIBRA annual modulation result has been pushed, being possible to reach five standard deviations by extending the data taking a few more years than the accumulated five years in August 2022.
\end{itemize}

In summary, the direct detection of DM particles is really challenging due to the small and rare signal expected and is being attempted by complementary experiments based on different detection technologies and targets and exploring different interactions, mass ranges of candidate particles and possible signatures. Multivariate machine-learning techniques are being applied in this field mainly to improve discrimination capabilities between signal-like events and different types of background or noise events.

\subsection{Muon identification and gamma/hadron discrimination using compact single-layered water Cherenkov detectors powered by ML techniques}

\subsubsection{Introduction: indirect detection of gamma rays with Extensive Air Shower arrays}
Very high-energy (VHE) gamma rays, ranging from 100 GeV to a few hundred TeV, can be used to investigate some of the most extreme non-thermal events taking place in the Universe, such as Active Galactic Nuclei (AGNs) and gamma-ray bursts (GRBs)~\cite{IceCube18,Abbott17a,Abbott17b}. Their indirect detection with a wide field-of-view (FoV) and a high-duty cycle is possible using Extensive Air Shower (EAS) arrays placed at high altitudes~\cite{albert2019science}. Observatories of this kind use dense arrays of detectors to observe the EAS of particles produced when gamma rays interact with the Earth's atmosphere. However, distinguishing the EAS induced by gamma rays from those by the vast cosmic-ray background is a challenge. Water Cherenkov detectors (WCDs) have proven to be effective for these purposes, as they allow having a duty cycle near to $100\%$ and detect muons, which are much more prone to appear in hadronic showers. Currently, despite the potential for mapping of large-scale emissions as the Fermi bubbles in the centre of our galaxy~\cite{FermiBubblesDiscovery,LATTES}, no such wide FoV experiment is operating in the Southern Hemisphere. Such an observatory would be complementary to the major facility CTA-South~\cite{acharya2013CTA} and enable full-sky coverage for transient and variable multi-wavelength and multi-messenger phenomena at this energy range. 

In this work, it is discussed the use of an EAS array composed of single-layered WCDs with multiple photomultipliers tubes (PMTs) for such purpose. By analysing the signals from the PMTs with a CNN, we aim to identify muons in the stations, and thereby discriminate between gamma-induced and hadron-induced showers in subsequent analyses.

\subsubsection{WCD concepts}

\begin{figure*}[t]
 \centering
  \subfloat[WCD design with three PMTs.]{
   \label{fig:MuonID:WCD_3PMTs}
    \includegraphics[width=0.35\textwidth]{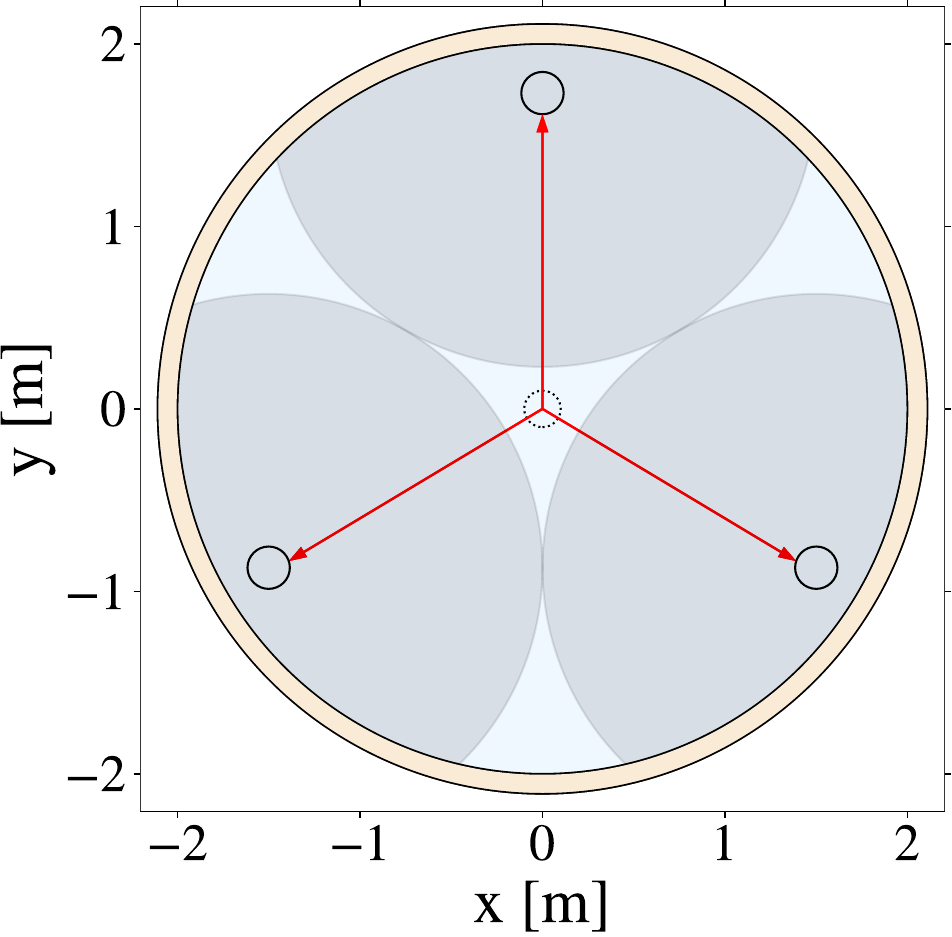}}
\hspace{0.5in}
  \subfloat[WCD design with four PMTs.]{
   \label{fig:MuonID:WCD_4PMTs}
    \includegraphics[width=0.36\textwidth]{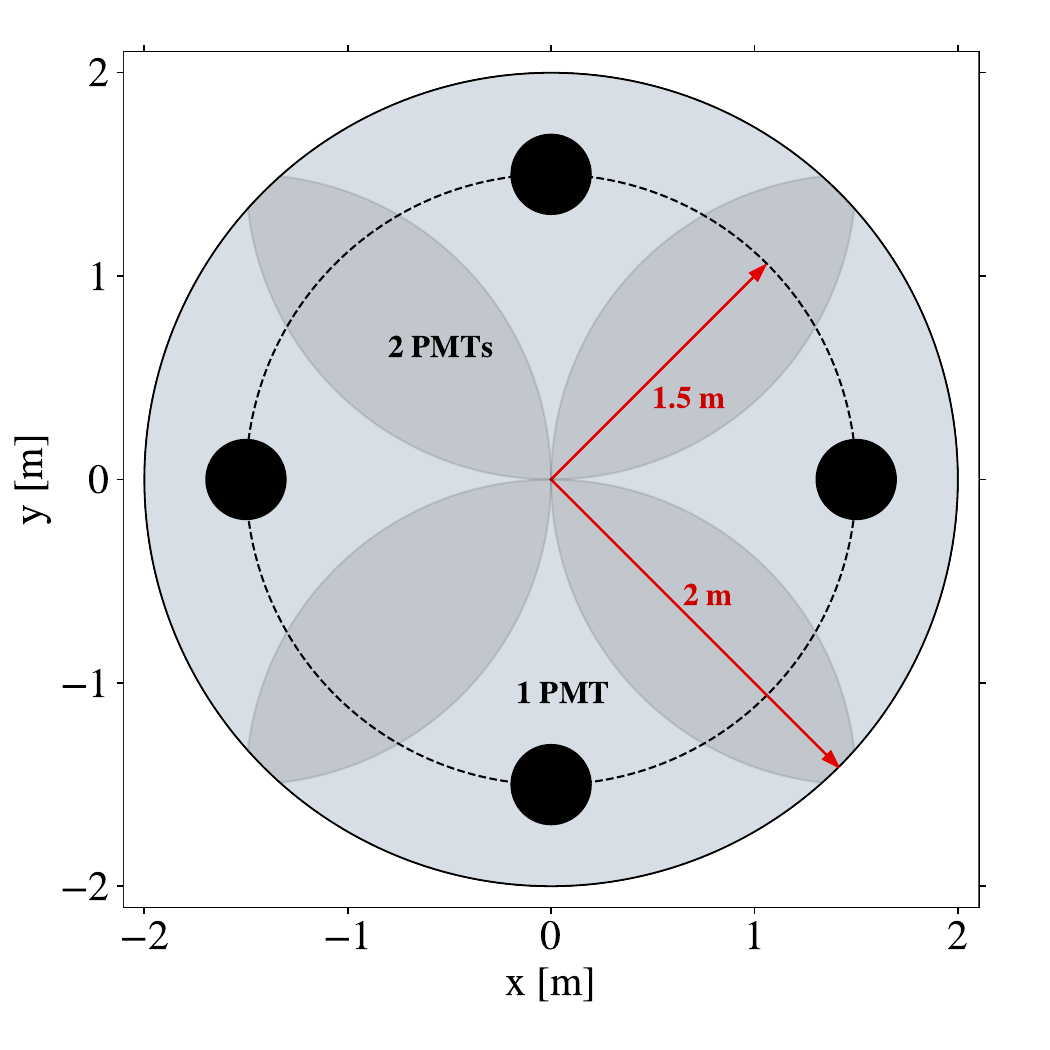}}
 \caption{Scheme of the single-layered WCDs. A dark circle with a 1.5m radius around each of the eight-inch PMTs was drawn. If a muon travels vertically through one of these circles, the direct Cherenkov light should be picked up by the corresponding PMT.}
 \label{fig:MuonID:WCD_detector}
\end{figure*}

The rationale behind the design of the WCDs used in this work is that muons will cross the whole detector, creating a direct Cherenkov light that reaches only a portion of the WCD floor, while photons and electrons will produce electromagnetic showers inside the station, creating a broader Cherenkov light pool. This asymmetry in the signal provides a mean to distinguish muons from other particles. 
The Cherenkov angle of a relativistic muon with an energy of $\sim$2 GeV in water is approximately $41^{\circ}$. Thus, to ensure complete signal coverage and a maximal signal asymmetry caused by vertical muons, the WCDs have a base diameter of 4m and a water height of 1.7m (see Fig.~\ref{fig:MuonID:WCD_detector}). 

The number of PMTs, and thus the dimension of the detector, can be optimised to balance the physics performance and the cost of the detector.
This design evolved from four (Fig.~\ref{fig:MuonID:WCD_4PMTs}) to three PMTs (Fig.~\ref{fig:MuonID:WCD_3PMTs}), as the distance of PMTs to the center is adjusted to optimise the detector while minimising the overlap of the areas covered PMTs, as well as the number of PMTs used. Both station concepts use eight-inches PMTs and white diffusive walls to maximise signal collection, which is essential to lower the energy threshold of the experiment.

This approach offers a cost-effective alternative to other methods that may increase the cost of the project. For example, the HAWC experiment~\cite{HAWC_GH} uses WCDs with a large volume of water and black walls to identify muons. Alternatively, other dedicated muon detectors can be buried or shielded as it is being done in the LHAASO experiment~\cite{LHAASO_muon} or placed below the WCDs as in the MARTA project~\cite{MARTA}.

\subsubsection{Simulation framework and sets}

The data sets used in this study were created using a simulation framework that combines the use of CORSIKA (v7.5600)~\cite{CORSIKA} for simulating extensive air showers and the \geant toolkit (v4.10.05.p01)~\cite{GEANT,allison_geant4_2006,allison_recent_2016} for simulating the detector's response.

The array configuration uses a dense layout, comprising $5\,720$ WCDs, covering an area of about $\sim 80\,000 \ \rm m^2$ with a fill factor of $85 \%$. The experiment observation level was set at $5\,200$m above sea level, which corresponds to the altitude of the ALMA site in Chile.

The gamma-induced showers were generated with energies $E_0$ in the range 1--1.6 TeV and a zenith angle of $\theta_0=10^{\circ}$, while the proton-induced showers were simulated with $E_0$ in the range 0.6--6 TeV and a zenith angle in the range 5--15$^{\circ}$. For both primary particles, the showers were generated following a $E_0^{-1}$ spectra (the events are afterwards weighted to ensure a realistic power-law spectrum of energies), and the azimuth angle was uniformly distributed. For each detector concept, more than $3\,000$ events were generated for gamma-induced showers, while for proton primaries more than $17\,000$ showers were simulated. 

Finally, a threshold on the total measured WCD signal at the ground is introduced to emulate a typical energy reconstruction, keeping events whose total signal at the ground is within one sigma around the mean of the gamma events.

\subsubsection{Single station performance}
The objective of this study is to determine the probability, $P_{\mu}^{(i)} \in \left[ 0, 1 \right]$, that a muon has passed through a given WCD with index $i$ based on the signal it has recorded. To achieve this, we use variables derived from the WCD signal time traces, as previously proposed in other studies~\cite{MuonID_NCA,MuonID_9SiPMs}. These variables aim to explore both temporal (patterns in the signal time traces) and spatial (asymmetry in the PMTs' integrals) features. The used variables include:
\begin{itemize}
    \item normalised signal time trace of each PMT;
    \item integral of each PMTs signal time trace;
    \item sum of the PMTs' signal trace integrals;
    \item normalised integral of each PMTs signal time trace.
\end{itemize}

The PMT signal is normalised to the sum of the signals recorded by all PMTs during the time window considered for the variable that will be normalised. Note that the normalised signal traces contain the first 30 nanoseconds to explore features from both direct Cherenkov light and the reflections, while the rest of the variables considered only the direct Cherenkov light (first 10 nanoseconds). Given the typical signal for vertical muons, only stations with more than 200 p.e. for the station with three PMTs and 300 p.e. for the station with four PMTs are considered.

A CNN was used to extract complex features from the signal time traces of the PMTs and combine them with the spatial features (integrals). With it, we define a regression problem to provide the probability $P_{\mu}^{(i)}$ that a muon has passed through the station. The algorithm's configuration was optimised using a validation data set for vertical showers. The network contains three convolutional layers to study the signal time traces and use ReLU activation function. The input of the first convolutional layer is set as one channel of data for each normalised PMT signal time trace. Afterwards, three dense layers are introduced to perform the regression using the previous signal features and the spatial variables. The model was trained with the Adam optimizer~\cite{kingma2014adam} during 200 epochs with a batch size of 512 and a learning rate of $10^{-3}$. Python 3.7 and Keras~\cite{chollet2015keras} were used as the framework of the entire study. A discussion of the CNN configuration and the use of various other approaches to this problem can be found in Ref.~\cite{MuonID_NCA}.

\begin{figure}[!t]
  \centering
 \includegraphics[width=0.45\textwidth]{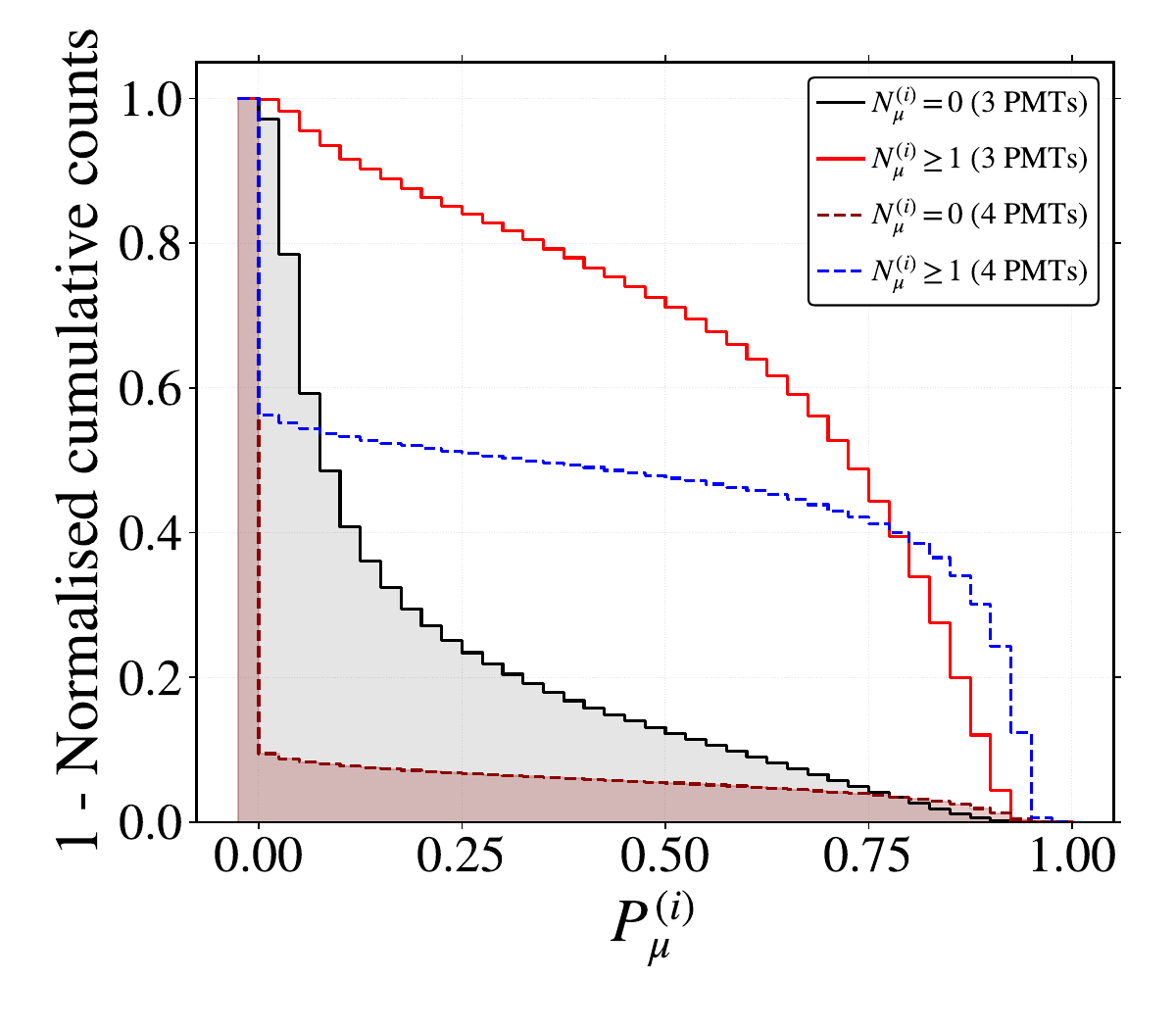}
  \caption{Inverse cumulative function of the $P_{\mu}^{(i)}$ variable for stations with and without muons using both detector concepts: three PMTs (full lines) and four PMTs (dashed lines).
}
\label{fig:MuondID:Pmui}
\end{figure}

In Fig.~\ref{fig:MuondID:Pmui}, the normalised inverse cumulative function for the probability $P_{\mu}^{(i)}$ is shown for both station concepts using proton-induced showers. The results were separated into stations with and without muons, being possible in the first case to have stations with both muons and electromagnetic contamination. Most stations with muons were correctly identified by requiring a probability $P_{\mu}^{(i)} \geq 0.5$. Roughly 70 and 50\% of the events passed the threshold for stations with three PMTs and four PMTs, respectively. 
Besides that, a low false positive rate was found for stations without muons, with approximately 15 and 10\% of them having a probability $P_{\mu}^{(i)} < 0.5$ for stations with three PMTs and four PMTs, respectively. A similar muon tagging capability can be achieved for inclined showers with an incident angle $\theta_0 \sim 30^{\circ}$, as long as the CNN is retrained for such events~\cite{Borja4PMTs}. 

\subsubsection{Gamma/hadron discrimination}
In this section, we describe a gamma/hadron discrimination strategy that relies on the use of the muon information extracted in the previous section at the single detector level, $P_\mu^{(i)}$. This information is combined at the shower event level to create a simple and intuitive observable, $P_{\gamma h}^{\alpha}$, which is defined as the sum of the probabilities $P_{\mu}^{(i)}$ to the power $\alpha$ as proposed in a previous study \cite{2021pgamma}.  Hence, we introduce the following discriminator quantity:

\begin{equation} \label{eq:MuonID:Pgh}
P_{\gamma h}^{\alpha} = \sum_{i=1}^{N_S} P_{\mu}^{(i) \alpha} (r_i \ \geq 40 \  \rm m)\,,
\end{equation}

where $N_S$ denotes the total number of WCD stations passing the threshold on the signal in the shower event. In addition to the threshold on the station signal, only stations far away from the shower core are sampled, to avoid the bulk electromagnetic shower component, which is much higher near the shower core. 

As described in Ref.~\cite{2021pgamma}, the introduction of the $\alpha$ power in the discriminator is motivated to reduce the relative importance of the large number of stations without muons in gamma and hadron high-energy showers ($E_0 \geq 40 $ TeV). In this work, since the relative importance of the number of stations without muons is not as high at $\sim$ 1 TeV energies, an $\alpha$ value of 1 is used.

This observable was tested for proton and gamma-induced showers with an equivalent total signal at the ground. It was possible to efficiently separate the gamma-induced showers from the ones generated by protons by adequately choosing a threshold on $P_{\gamma h}^{\alpha}$.

\begin{figure}[!t]
  \centering
 \includegraphics[width=0.45\textwidth]{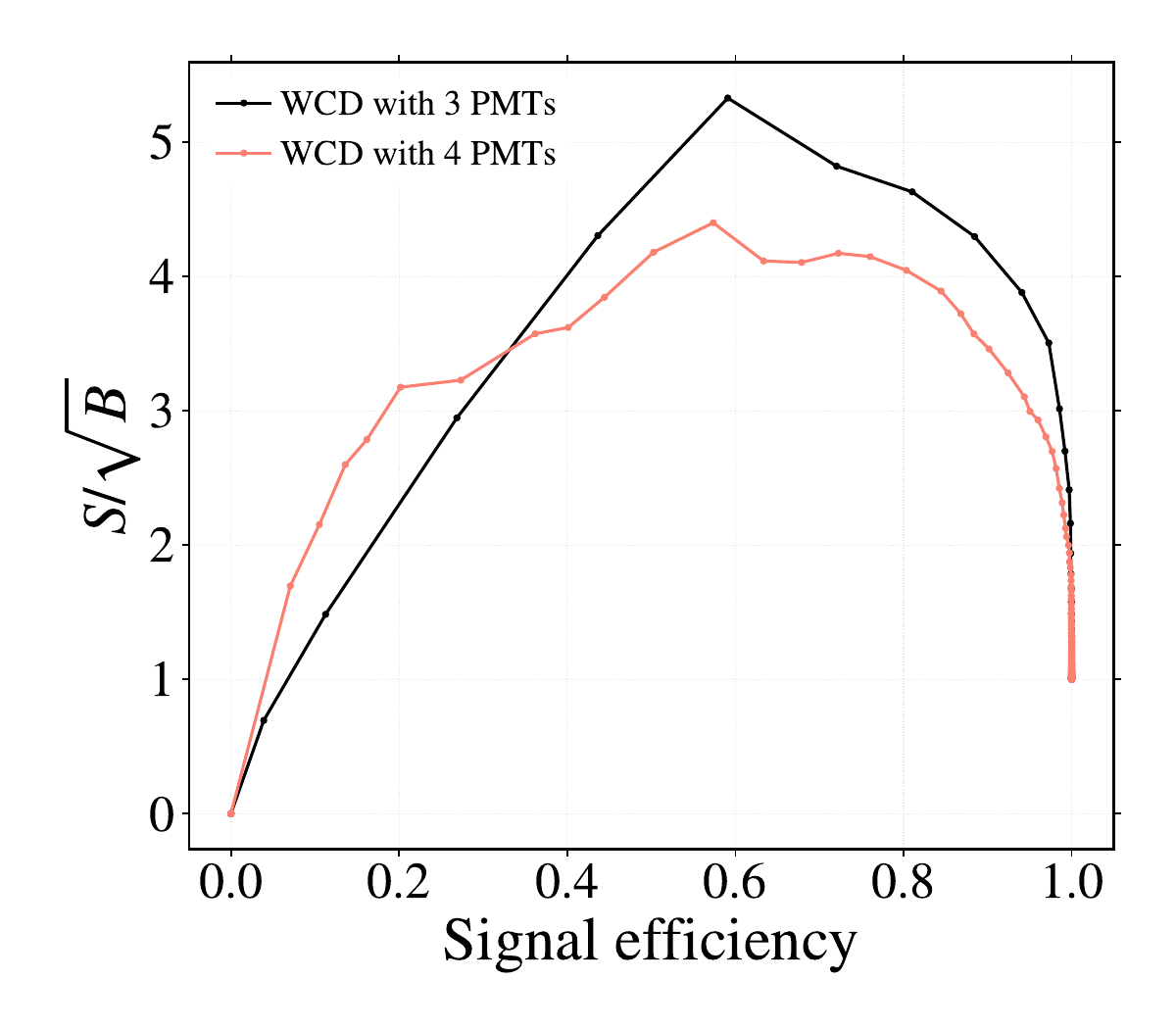}
  \caption{Ratio of the selection efficiency of the discriminator $P_{\gamma h}^{\alpha}$ for gammas ($S$) over the square root of the selection efficiency for protons ($B$) as a function of $S$. The dashed line corresponds to the result obtained with four PMTs \cite{Borja4PMTs}, while the full line corresponds to the results obtained with three PMTs~\cite{wcd2022mercedes}.
}
\label{fig:MuonID:GH_eff} 
\end{figure}

A proxy to a gamma-ray experiment flux sensitivity can be obtained by evaluating $S/\sqrt{B}$, where $S$ and $B$ are the selection efficiency for gamma rays and the background (protons induced events), respectively. As shown in Fig.~\ref{fig:MuonID:GH_eff}, a $S/\sqrt{B} \sim 4$ was found for both detector concepts when fixing the selection efficiency for gammas to $S=0.8$. The obtained value is similar to the one quoted in other experiments such as HAWC~\cite{HAWC_GH} and LATTES~\cite{LATTES}, but using significantly smaller WCDs than HAWC.

\subsubsection{Conclusions}
In this section, we have shown that a WCD with a reduced water volume of $12\,{\rm m^2} \times 1.7\,{\rm m}$ and multiple PMTs placed at the bottom can be used to efficiently tag muons. A ML-based analysis was developed to process the PMT-acquired signals. 
The results showed that the identification of muons in stations, given as a probability, $P_\mu^{(i)} \in \left[ 0;1 \right]$, can be used to build a simple gamma/hadron discrimination observable able to distinguish between gamma and proton-induced showers with a $S/\sqrt{B} \sim 4$ for shower energies of $E_0 \sim 1$ TeV and $\theta_0 \sim 10^\circ$ and a similar total signal at the ground.

\subsection {End-to-end optimization of the SWGO array layout}

The Southern Wide-Field Gamma Observatory (SWGO) is planned to be built at high altitude in south America (possible sites being considered for its construction are in Peru, Chile, and Argentina). The observatory will consist of several thousand Cherenkov detectors in the guise of PMT-endowed water tanks, deployed in an array spanning a footprint of up to a few square kilometers; alternatives include the deployment of photodetectors in excavated pools or in submerged bladders on the surface of a lake. The scientific interest of such a detector, which will be sensitive to primary cosmic gamma rays in the multi-GeV to multi-PeV range, lays in its capability of studying a number of sources in the southern sky, with continuous operation, excellent pointing and energy resolution, and very high discrimination of background from hadronic primaries. 

At the time of writing, a number of different configurations for the precise design of the water tanks are being considered. They share the need to count both the collective number of electrons, positrons and photons produced in the electromagnetic shower development, as well as the number of muons, which are very rare in gamma ray-originated showers and thus constitute a powerful discriminator of hadron backgrounds. Due to the higher penetration power of muons, Cherenkov tanks may be able to distinguish these particles by the shape of the light signals collected by properly placed photomultiplier tubes on the bottom of the tank, or by having PMTs pointing both upwards and downwards in the middle of the tank. Optimization studies of these designs are ongoing, with the aim of finding the right compromise between cost, logistics, and performance of these units.

A separate issue concerns the arrangement of the tanks on the ground. The footprint of energetic air showers may span several hundred meters or even kilometers at the height above sea level of SWGO, and the capability of the array to appropriately measure the energy of the primary particle in the low-energy range (tens-hundreds of GeV) requires the collection of as large a fraction of the collective signal of particles on the ground as possible; this mandates that tanks be laid out as tightly packed as possible---\ie with a "fill factor" (in terms of fraction of ground covered by the active area of the detectors) of 50\% or more. On the other hand, in order to be sensitive to the very low flux of photons of extremely high energy (1 PeV and above), the total instrumented area (including the void between units) needs to be maximized with detectors distributed with a low fill factor, well spaced from one another. However, too low fill factors will hinder both a precise energy reconstruction and a successful separation of hadron backgrounds, whose rate at those high energies exceeds that of gamma rays by a factor exceeding $10^4$. 

The above contrasting requirements have led the SWGO collaboration to propose a set of different layouts that share as a common feature a dense core and an extended region more sparsely populated. Investigations are ongoing to evaluate in quantitative terms the relative benefit of the proposed layouts for the various astrophysics cases of interest to the collaboration. The more principled question however concerns the investigation of the way more high-dimensional space of all possible configurations of N detectors. As a detector position is specified by its ground coordinates $x,y$, the space of configurations lives in ${\mathbb{R}}^{2N-3}$, accounting for the azimuthal symmetry of the problem. Therefore, \eg, the layout of three units involves the choice of three real parameters: by setting the first unit in (0,0), the second can be set at ($x_2$,0) without loss of generality; the third can then be specified by ($x_3$, $y_3$), when $y_3$ can be chosen in the positive semi-axis without loss of generality. With over 6000 units to be deployed, the space of possibilities is thus prohibitively large to be probed with discrete methods.

The problem of writing an optimization pipeline that scans the large parameter space with differentiable programming can be addressed only if we endow ourselves with a parametrized approximation of the density of particles on the ground as a function of the distance from the shower axis, resulting from a cosmic ray shower of given energy $E$ and polar angle $\theta$. Ideally this should be available in closed form for both gamma and proton-originated showers (the background to discriminate against), and for the particle types to which the detectors are differently sensitive: electrons, positrons, and photons (which can be in first approximation be considered together, due to their similar behaviour in interacting with water at energies much above the critical energy $E_{crit}$, which is of 76.2 MeV for positrons and 78.3 MeV for electrons~\cite{Workman:2022ynf}), and muons. As atmospheric showers arising from high-energy primaries involve a number of complex stochastic processes, precise parametrizations are difficult to produce; yet even an imprecise choice may be able to successfully inform a continuous scan of the parameter space, if a full simulation is then used to validate the results in the vicinity of interesting configuration points.

We used large CORSIKA~\cite{CORSIKA} simulations of gamma and proton showers at different energies (from 100 GeV to 10 PeV) to extract a model of the density of secondary particles on the ground as a function of the distance from shower axis. These were the basis of an optimization pipeline, which we now describe. 

\begin{enumerate}
    \item The starting point is the generation of an initial ground configuration of N detector units,  
    ${(x_i,y_i)}_{i=1,...N}$. 
    \item A set of gamma and proton showers are simulated with an intersection of the shower axis with the ground, at a position $x_0,y_0$ randomly chosen within a region suitably exceeding on all sides the footprint chosen for the detector tanks, such that showers at the edge of this region have a negligible probability of being detected by the array. Given shower energy $E$ and polar angle $\theta$, the position of the shower center on the ground, together with the azimuthal angle $\phi_0$ defined as the angle between the shower axis and the positive $x$ direction, determines an estimate of the average number of particles of different kinds that will hit the sensitive area of each of the N detector units. These numbers can be sampled from Poisson distributions of means equal to the estimated averages.
    \item Through a likelihood maximization, estimates of the shower parameters are obtained by considering the true value of the parameters of the model, and the number of particles detected for each species in each tank under, for the time being, the assumption that tanks have 100\% detection efficiency and perfect discrimination power between the various particle species. The likelihood is maximized under both hypotheses (gamma or proton) for each shower, regardless of the true primary particle species.
    \item The obtained maximum likelihood values, maximized over shower parameters E, $\theta$, $\phi$, $X_0$, and $Y_0$ are used to compute a likelihood ratio test statistic $T$ for each generated shower. The sampling of a large number of showers of different parameters allows to construct a PDF of $T$ for both hypotheses.
    \item A new batch of gamma and proton showers is generated, and a distribution of $T$ values obtained. This is fit as the sum of the two distributions, and the uncertainty of the fraction of gammas in the batch is extracted as the Rao-Cramer-Frechet bound. 
    \item Using the uncertainty on the gamma fraction, a utility function is computed as the weighted sum, for different energy points, of the inverse of the relative uncertainty in the flux of gammas in the batch. 
    \item A propagation of derivatives then allows to extract the derivative of the utility function over displacements $\delta x$, $\delta y$ for each of the N detector units. These are used to update the detector positions.
    \item The cycle can be continued by generating a new set of gamma and proton showers and deriving new PDF of T, then fitting a batch of showers and recomputing the utility function $U$. 
\end{enumerate}

The pipeline described above allows to converge to layouts that optimize a simple utility function focusing on the precision of the gamma flux. Of course, more complex utility functions could be conceived, to model the real objectives of the wide scientific program of the SWGO collaboration. This modeling task has not yet been attempted. In the near future the code will be expanded to improve the way it represent the full inference-extraction procedures, with a view to providing useful input to the collaboration on the most advantageous layouts of tanks on the ground. The relative gains of those layouts over baseline ones can then be appraised by exploiting full simulation of atmospheric showers and detection of secondaries by the detector units.

In parallel, after a final decision is taken on the general design of the detector units, a parametrization of the detection of particles of different species may be included in the pipeline, improving the precision of the model.

\clearpage
\section{Progress in Neuromorphic Computing Optimization}
\label{sec:neuro}
Neuromorphic computing (NC) is an emerging computing paradigm that exploits brain-inspired, time-encoded signal processing to allow for highly energy-efficient, highly parallelizable, and decentralized information processing including fast inference of data-driven models~\cite{mead2020we,christensen2022roadmap}. In addition to other novel models of computation, such as quantum computing, NC promises a major path for beyond-Moore computing and artificial intelligence. Unlike quantum computing, NC relies on less exotic hardware and focuses on designing electronic systems, such as processors and memory devices, that are inspired by the structure and function of biological neural networks.

The primary motivation for this approach is to overcome the limitations of conventional computing technologies, particularly in cognitive tasks such as low-power perception and learning, which the human brain excels at. By leveraging biomimetic principles, NC systems offer the potential for real-time processing, low power consumption, and enhanced efficiency, paving the way for the next generation of intelligent machines and technologies that can potentially transcend Moore's law~\cite{shalf2020future,leiserson2020there,mehonic2022brain}.

Below we provide a brief introduction and an overview of the status of NC and its applications at large, after which we discuss a few ideas of potential exploitation in fundamental science of the specific functionalities and prerogatives that NC offers. 

\subsection{Neuromorphic computing introduction}

It is helpful to recap some basic neuroscience concepts~\cite{gerstner2014neuronal} in order to understand how NC is different from conventional deep neural networks and related hardware accelerators.
Biological neurons have ion pumps that maintain a Nernst potential of about $-70$~mV
with respect to the the exterior of the cell membrane.
The time-varying potential depends on the flow of ions (Na$^+$, K$^+$, Cl$^-$, etc) through channels in the cell membrane, which open and close depending on the electric field gradient and concentration of neurotransmitter molecules etc.
Neurons have rich temporal dynamics, signaling patterns, and adaptation mechanisms at different temporal and spatial scales that mediate processing and storage of information. AD-powered differentiable models of neuronal dynamics are compared with data from electric fishes in Ref.~\cite{vischia_pietro_2023_8394819}.
Neuromorphic devices mimic these processes either at the physical level through mixed-signal electronic circuits, which for example emulate the ion diffusion processes in neurons with energy-efficient analog CMOS circuits operating in the subthreshold regime, or through custom digital systems offering flexible simulation capabilities.

Due to a high variability of biological computation mechanisms evolved naturally, there is a remarkable diversity of approaches falling into the bucket of NC, such as:

\begin{itemize}
\item \emph{Memristive Systems}: Memristive devices are a class of electronic components that exhibit memory and resistive switching characteristics. These properties enable them to mimic synaptic and neuronal dynamics, providing an alternative way to design neural networks and learning in neuromorphic systems~\cite{payvand2022memristors}.
\item \emph{Dendritic Integration}: Neurons are observed to perform a variety of different forms of synaptic integration on their inputs, emphasizing the critical role of dendrites in the brain and for understanding how neurons learn and integrate the thousands of synaptic inputs. The central role of dendrites suggest that in the search for more energy-efficient NC solutions we should move beyond learning with synapses to learning with dendrites~\cite{boahen2022dendrocentric}.
\item \emph{Recurrent Networks}: This approach includes large, randomly connected recurrent neural networks known as 'reservoirs' with nonlinear recurrent dynamics that act as a temporal kernel for the input signals, which can be further processed/classified using a trained readout network. This research domain includes early approaches such as Liquid State Machines (LSM).
Recurrent networks are particularly relevant for tasks involving time series data and pattern recognition~\cite{rao2022spikelstm, zhu2023spikegpt}.
\item \emph{Brain-inspired Cognitive Architectures}: These entail developing microelectronic circuits, high-level modular architectures and algorithms that capture the cognitive features of the brain, such as memory, perception, learning, and problem-solving~\cite{bartolozzi2022cognitive}.
\item \emph{Efficient algorithms}: NC opens new algorithmic opportunities for efficient solutions to some conventional hard computational problems such as constrained optimization, graph algorithms, kernels for composition, and signal processing~\cite{aimone2022review}.
\end{itemize}

While each of these directions possesses unique strengths and limitations, they collectively contribute to the progress and diversity of NC research, offering a variety of paths towards emulating the computational efficiency and cognitive capacity of the brain.
For reviews, see Refs.~\cite{christensen2022roadmap, schuman2022opportunities, aimone2022review, mehonic2022brain} and references therein.
In the following we focus on some general aspects of NC architectures, which can influence the future of fundamental science computing systems and design optimisation solutions.

\subsection{Spiking Neural Networks}

\begin{figure}[tb]
    \begin{center}
        \includegraphics[width=0.99\textwidth]{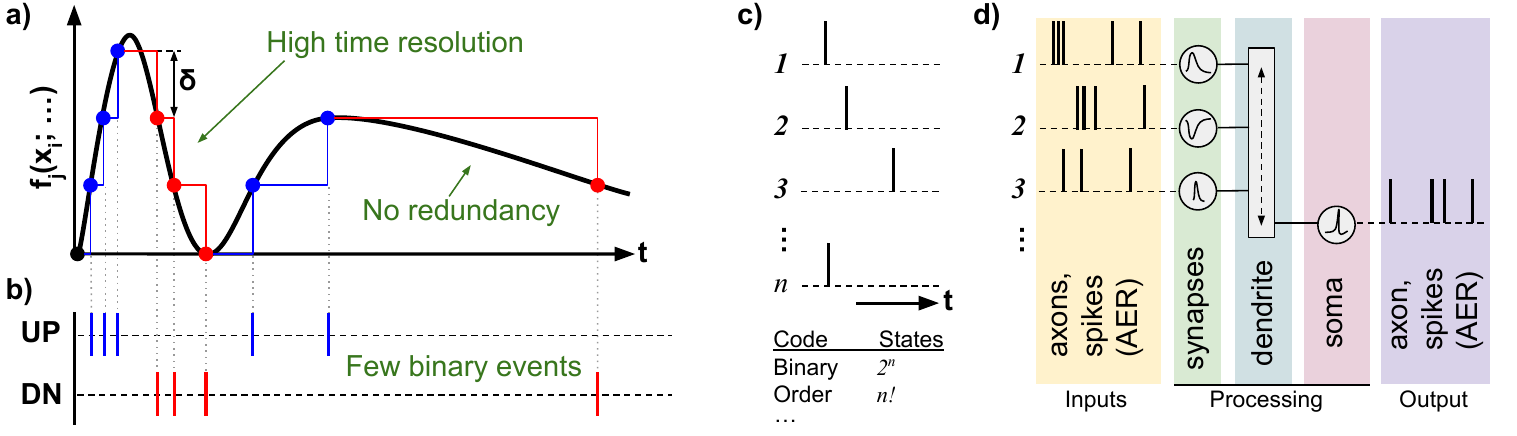}
        \caption{Illustration of (a) Lebesgue sampling of a signal, $x_i$, transduced from the environment, (b) the resulting signal encoding as unary {\it spikes} on two channels representing changes of $+\delta$ (UP) and $-\delta$ (DN), (c) spike-based encoding of information across $n$ channels representing axons, and (d) a schematic of a spiking neural network unit with input and output channels. In general, the performance of a system comprising such parts is subject to optimization of the sensors, $f_j(x_i,\ldots)$, sampling parameters such as $\delta$ and noise shielding mechanism, spiking neural network hyperparameters, and hardware realizations for the application at hand.}
        \label{fig:neuromorphic:spikecoding}
    \end{center}
\end{figure}

A Spiking Neural Network (SNN) aims to mimic the behavior of biological neurons and neural networks more closely than conventional artificial neural networks.
In SNNs, neurons communicate using time-encoded events called \emph{spikes}, similar to how neurons in our brain encode and transmit information using (mostly stereotype) electrical pulses in the form of action potentials.
Spikes are asynchronous, unary ``one-or-nothing'' events where the physical timing of the spikes in relation to the dynamic states of neuron units and the environment encode information, as illustrated in Fig.~\ref{fig:neuromorphic:spikecoding}.
Time-based signal encoding is a key characteristic that differentiates SNNs from traditional artificial neural networks, making them more biologically plausible and in some cases more resource-efficient information processors. Instead of sampling and processing signal amplitudes at some time interval (the conventional Nyquist/Riemann paradigm) each sensor channel asynchronously generates one bit of information if and only if the corresponding signal feature has changed by some amount (Lebesgue sampling, Fig.~\ref{fig:neuromorphic:spikecoding}a). This way input signals are transformed into patterns of precisely timed spikes, with the information being encoded in the relative timing of the spikes (Fig.~\ref{fig:neuromorphic:spikecoding}b). Level-crossing analog to digital converters are basic sampling devices of this type. Spike-based representations of information allow SNNs to process and propagate information efficiently across the network: see Ref.~\cite{thorpe2001coding} for an introduction. For example, in comparison with a synchronous binary encoding on $n$ channels an asynchronous spike-ordering code can represent $\log_2(n!)$ bits of information given sufficient timing precision (Fig.~\ref{fig:neuromorphic:spikecoding}c). Synapse conductances, dendritic processes and neuron voltages are modeled with time- and sometimes also space-dependent differential equations under the assumption that action potentials can be approximated with spikes that are characterized by the spike time (Fig.~\ref{fig:neuromorphic:spikecoding}d). In neuromorphic systems, spikes are typically encoded and communicated using address event representation (AER) to improve efficiency, where the spike time is represented by physical time while only the source neuron address has an explicit binary representation. Given an SNN neuron connectivity table the AER spike representation includes the minimum binary information needed to determine the receiving neuron addresses.

There are no graded activation functions in SNNs (such as ReLU, sigmoid etc.), but the mean spiking frequency corresponds in a quasistationary sense to the activation of an artificial neural network (ANN) unit. Thus, ANNs resemble ``mean-field models'' in the sense that they model mean firing rates of neurons, hence the commonly used term ``rate-based'' neural networks, while SNNs model the time-dependence of neuron potentials (with varying spatial resolution through point-like, multi-compartment or finite-element models) under the assumption that action potentials can be represented with spikes. This spiking mechanism introduces an additional layer of complexity and dynamics to the network, allowing it to efficiently encode and process temporal information. SNNs are therefore well-suited for tasks requiring real-time processing and adaptation, such as sensory processing, motor control, and decision-making. As a result, they have become a major focus in the field of NC, offering the potential to unlock new paradigms in artificial intelligence and algorithm development.

\subsection{Neuromorphic computing optimization challenges}

With DL computing needs doubling every few months the foundational optimization algorithms and digital technology drive the energy and resource requirements of artificial intelligence at an unsustainable rate~\cite{mehonic2022brain,thompson2021deep}, making this vital technology inaccessible for broad and safe societal benefit. The development of neuromorphic technologies provide hints to what may become the future of ubiquitous computing, optimization and artificial intelligence.

In principle it is possible to perform end-to-end optimization of SNNs using surrogate gradients for the non-differentiable parts~\cite{eshraghian2023training}. However, training of SNNs using state-of-the-art back-propagation through time (BPTT) is costly and can lead to numerical instabilities if not handled carefully~\cite{bauer2023exodus}. Furthermore, the task of systematically optimizing the design of a system, including sensors, spike based encoding, spiking/artificial neural network modules, and the cognitive architecture given specific actuators and application constraints is an open challenge. This is an opportunity for research and development in NC design optimization using differential and probabilistic programming. Progress in this direction can enable valuable advances in the co-design and optimization of the hardware--software--algorithm stack~\cite{schuman2022opportunities} and edge-to-cloud computing continuum~\cite{nilsson2023integration}. Furthermore, considering the knowledge gaps and broad interdisciplinary NC efforts to understand and mimic information processing and optimization in the brain, bridging these communities can lead to radically new insights in resource-efficient optimization and design of experiments and systems in general.

Event-driven architectures can significantly reduce the average power of sensor systems that operate in random-sparse-event physical environments, for example in the form of an always-on wake-up circuit that triggers a conventional high-performance computing system. Note that this is also possible when the relevant frequencies in the environment are much higher than the spike rates and time scales in the SNN domain~\cite{adam2020reconstruction}. However, there are also challenges related to noise management and precision, see for example Refs.~\cite{ye2021challenges,safa2023lcadcopt}. In the presence of noise, a level-crossing sampler of the type illustrated in Fig.~\ref{fig:neuromorphic:spikecoding}a-b generates false events that waste energy and downstream processing performance. Different methods have been investigated to overcome this difficulty, such as introducing a hysteresis window around each amplitude level or a noise-shielding time window after each sample. When high amplitude precision is required a conventional uniform-sampling ADC is more efficient because the number of level-crossing samples tend to increase exponentially with resolution, while a level-crossing sampler can offer relatively high dynamic range and temporal resolution. Further work is required to enable a systematic approach to optimizing the design of systems incorporating both Nyquist/Riemann and Lebesgue sensor degrees of freedom, as well as hybrid SNN-ANN cognitive architecture and NC hardware co-design optimization.

\subsection { Spiking neural networks for low-momentum tracks removal in CMS pixel layers}
Silicon pixel detectors allow for precise measurements of charged particle tracks and vertices at collider experiments. Next-generation detectors will require a reduction in pixel size, leading to unprecedented data rates exceeding those foreseen at the HL-LHC. Signal processing that performs a physics-motivated filtering of the data within the pixelated region of the detector \textit{at the collision rate} will enhance physics performance in a high luminosity environment.

The shape and time evolution of charge clusters deposited in an array of small pixels encodes information about the physical properties of the traversing particle.  These can be extracted with locally customized neural networks implemented directly in the front-end electronics. A first demonstration takes the form of a filtering algorithm that rejects tracks with transverse momentum $p_T$ below the threshold that is useful for physics analysis. 

Spiking neural network models offer several advantages for this machine learning task: the time evolution of the charge cluster is naturally encoded in the input data, and the minimal number of parameters and low power requirements are well suited for implementation on an ASIC. The incoming charge waveforms can be converted to streams of binary-valued events which are processed by the SNN. Studies show that an SNN trained using an evolutionary algorithm and with optimized set of hyperparameters shows a signal efficiency more than 90\% for for particles with $p_T>2$ GeV, using a factor of two fewer parameters than a similarly trained Deep Neural Network~\cite{kulkarni2023onsensor}.

\subsection{Integration of neuromorphic computing in online triggers at hadron collider experiments}

The data rate yielded by high energy physics colliders such as the LHC is of the order of the tens of megahertzs. Online triggers systems are responsible for
bringing the rate down to $\mathcal{O}(1\mathrm{kHz})$, a rate compatible with data storage systems. A recent study~\cite{10.1063/5.0116699} has demonstrated that it is possible to reduce the trigger rate from 40 MHz to about 75kHz trigger using neuromorphic chips when considering anomaly-detection-based triggers. In comparison to CPU and GPU benchmarks, neuromorphic chips offer throughputs larger by over a factor 20 for the largest networks that were tested. 

These results open the road to further testing and to the deployment of "Level-1" trigger menus implemented in neuromorphic chips.

\subsection{Integration of neuromorphic computing in fine-grained calorimeter readout} 

In parallel to the investigations of granular calorimetry discussed in Section~\ref{sec:calopt}, it appears interesting to study the possibility to integrate edge computing elements based on neuromorphic processors in the readout of the device. While the typical aim of NC are applications where extremely low power consumption and computing at the data-generating end are required, we speculate that the potential of localized preprocessing and extraction of information within the core of a highly granular calorimeter could be quite significant. A calorimeter made of millions of individual cells, whose complete readout would pose significant challenges, might strongly benefit from a localized, ultra-fast, and ultra-low-power pre-processing of the information at the highest spatial resolution, before higher-level primitives can be transferred to the back-end for reconstruction. In addition, the possible timing information that new generation detectors are starting to enable would find a perfect match with the time-encoded specific capabilities of timeconstant-modified NC hardware processors based on, e.g., state-of-the-art CMOS, nanowire memristor networks, and photonic and opto-electronic technologies .

The work plan in this case involves:
(1)	A demonstration of the unsupervised learning of specific patterns in macro-cells, via {\em in-situ} neuromorphic processing of full granularity information via emulation in digital processors;
(2)	Study of possible designs and hardware implementations that incorporate that functionality, and their effectiveness in terms of latency, information extraction, power consumption, and heat generation with respect to ordinary digital computing solutions;
(3)	Prototyping, if the solutions developed in (1) and (2) prove effective.

The lack of existing hardware solutions to the problem of increasing the processing speed of NC devices to a level suitable for fundamental science applications such as the ones we discussed in this section demands a wide-ranging parallel study of photonic and opto-electronic solutions for the emulation of neuronal decoding of signals which are natively encoded as light pulses in conceivable homogeneous calorimeter elements.

\clearpage
\section{Progress in Medical Physics}
\label{sec:medical}

In this Section we outline recent progress in medical physics made possible by AD/PD techniques.

\subsection{Fast emulation of deposited dose distributions by means of Graph Neural Networks}
Cancer is a leading cause of death worldwide, accounting for nearly 20 million cases and 10 million deaths only in 2020. Among the various existing therapies, one of the most effective and frequently used treatments is external beam radiotherapy (RT). 
For every RT treatment, an essential part of the process that precedes the effective delivery of radiation is the treatment plan optimisation. Such a phase consists in choosing therapeutic beams’ energies and fluencies as a function of their orientation in order to fit the medically prescribed dose, with particular attention on the dose that should be received by the tumor and the maximum deliverable dose to organs at risk (OARs). Currently, this optimization is done using reliable but standard sequential algorithms that optimize energies and fluencies in two different steps. Often, such algorithms suffer the need of reaching a trade-off between the optimality of the solution and the computing time and in most cases, they are not suited to handle new complex tasks in reasonable times.
Novel therapies, such as electron FLASH RT~\cite{schuler2017very} or Volumetric Modulated Arc Therapy (VMAT)~\cite{otto2008volumetric}, offer much more freedom in the choice of the entry angles of the beam. The dose can be shot to the patient from a wide set of angles, reducing significantly collateral damage to healthy tissues, while delivering the right amount of dose to the tumor. However, this possibility entails a significant increase in complexity in the optimization process which can lead to a significant increase in computing time. In principle, the dose should be computed and optimized for all the possible orientations of the beam accelerator, with continuity.

In this context, DL algorithms, based on new tools such as tensorization, GPU acceleration, and AD, can represent a way to overcome such limitations.
The aim of our study is to construct a DL model capable of generating deposited energy distribution data as a function of beam parameters and medium density in negligible time and with high precision. Such a model could represent the basis for a new generation of optimization algorithms.
At this stage, we developed a Variational AutoEncoder (VAE), based on Graph Neural Networks, that can generate deposited energy distributions inside a simplified voxelated material. 
In the following, we show a representative sample of the studies already published in Ref.~\cite{Arsini2023nearest}, where the full description of the DL model and some ablation studies are presented. Moreover, we will show some unpublished results on the VAE latent space and sample generation.

\subsubsection{Dataset}

The dataset employed in this work was built running simulations using \geant~\cite{GEANT,allison_geant4_2006,allison_recent_2016}, a toolkit in C++ for Monte Carlo simulations in particle physics.
We simulated electron beams, made up of 10\,000 primaries each, fired toward a voxelated material.
Such material consists of a water cube, with a side length of \SI{80}{\centi\meter}, containing a \SI{3.5}{\centi\meter} thick slab of with variable density, perpendicular to the beam.
For each run of the simulation, we sampled uniformly the particle energy between \SI{50} and \SI{100}{\mega\eV} and the slab density between \SI{0} and \SI{5}{\gram\per\centi\meter\cubed}.

Energy deposition data are collected in a cylindrical scorer, aligned with the electron beam, of \SI{50}{\centi\meter} height and \SI{5}{\centi\meter} radius, divided in $28 \times 28 \times 28$ voxels along the $z$, $\theta$, and $r$ axes. 
This choice brings two main advantages. First, without any loss of generalization, we are reducing the complexity of the generation task. We don't want the network to generate the deposited energy distribution in the whole volume, which may depend on beam orientation and, in real applications, on the organs' shapes and dimensions. We just consider a cylindrical volume around the beam, which still can be made large enough to contain all the energy released by the beam.
The second advantage is that with a cylindrical shape, we gain more precision near the beamline, where more energy is deposited and a precise prediction is most important.

The dataset is made up of 6\,239 examples of energy distributions and is divided into train, validation, and test set.
The test set represents a relevant central subset in the parameter space. It only contains examples with both particle energies ranging between \SI{70} and \SI{80}{\mega\eV} and slab densities ranging between \SI{2} and \SI{3}{\gram\per\centi\meter\cubed}, which account for the 4\% of total data samples.
These examples, are removed from the dataset and used for testing the network's ability to interpolate between samples.
The remaining data is used for training and validation.
In this way, we reduce the probability of generating artifacts in the model predictions at the edges of the test set.
In particular, we used $\sim5\,400$ examples for training and $\sim600$ examples for validation.

\subsubsection{Deep Learning model}

To emulate deposited energy distributions we employed the generative network described in Ref.~\cite{Arsini2023nearest}.
It is a VAE~\cite{Kingma2014auto} in which both encoding and decoding are based on Graph convolutional layers to fully take advantage of the cylindrical structure of our data.
A VAE is a generative model made up by two networks: the encoder and the decoder.
The encoder takes the input data and maps them to a distribution in a low-dimensional space, called latent space. The decoder is then trained to retrieve the input data starting from a point sampled from that distribution. The latent space is forced to be continuous; in this way, it is possible to generate new data with continuity, sampling from the latent space.

In our case, first, a suitable graph structure is imposed on the data. A node is associated to each voxel and nodes are connected to each other with nearest-neighbor connectivity. 
Graph data is then processed by the encoder which applies GraphConv layers~\cite{morris2019weisfeiler}, and lowers the graphs dimensionality using ReNN-Pool, a geometry-based pooling operation we developed for this task~\cite{Arsini2023nearest}.
The encoded data is then processed as in standard VAE architecture with Gaussian prior, using the reparameterization trick, mapping the input data to Gaussian distributions in the latent space, which, in this case, is set as two-dimensional.
The decoding uses the same graph representations employed for the encoding, but in reverse order. Nodes from a lower dimensional representation are up-sampled using the inverse operation of ReNN-Pool, and further processed by graph convolutions until the original graphs are recovered.
The model is trained using the standard VAE loss with binary cross-entropy as the reconstruction term, and the Kullback-Leibler divergence term.
We employed the Adam optimizer for weight updates, starting with a learning rate of 0.003. Additionally, we utilized an exponential scheduler with a decay factor of $\lambda = 0.9$.

\subsubsection{Results and discussion}
\paragraph{Reconstruction}
The DL model was trained for 200 epochs on a Tesla V100 SXM2 GPU and the best set of learnable parameters was chosen as the one that minimizes the validation loss. Training was stopped at 200 epochs observing that the validation loss ceased decreasing in the last 40 epochs.

\begin{figure}[h!]
\begin{center}
\includegraphics[width=.999\textwidth]{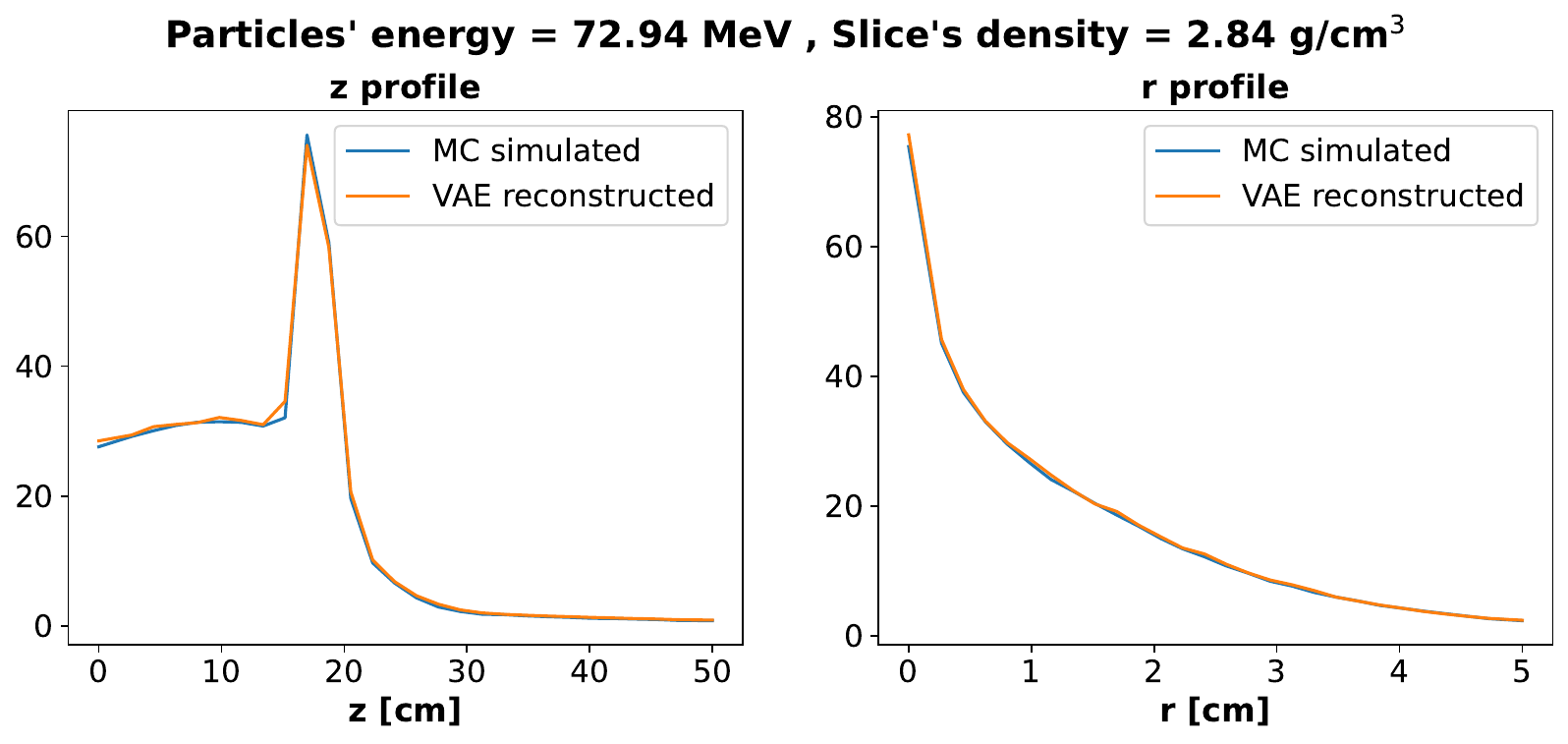}
\caption{\textbf{Energy profiles reconstruction}. Distribution of energy deposition along $z$ and $r$ axes. The blue line correspond to the Monte Carlo simulated data, while the orange line refers to the reconstructed data from our Network.}
\label{fig:dose_arsini:profiles}
\end{center}
\end{figure}

In Fig.~\ref{fig:dose_arsini:profiles} we show the energy profiles along the $z$ and $r$ axes, obtained by integrating the deposited energy distribution data along the other two dimensions, $r$, $\theta$ and $z$, $\theta$, respectively.
In each panel, the orange line corresponds to the reconstructed data from our Graph VAE, while the blue line refers to the ground truth, \ie Monte Carlo simulated data.
In Table~\ref{table} we report the mean relative errors on profiles and total energy deposition, with their standard deviation over the test set. Note that most of the errors are relative to the tails of the energy distribution, where the fluctuations of the Monte Carlo simulation are not always negligible. 
\begin{table}
  \caption{Mean relative errors on energy profiles and total energy on test set. values are reported along with their standard deviations on the test set.}
  \label{table}
  \centering
  \begin{tabular}{lrrr}
   \toprule
         & $Z$ profile     & $R$ profile & Total energy \\
    \midrule
    Relative error &  6.9 $\pm$ 3.4\% & 3.0 $\pm$ 1.2\% & 2.2 $\pm$ 1.6\%     \\
    \bottomrule
  \end{tabular}
\end{table}
To quantify the node-per-node reconstruction quality we employ the $\delta$ index~\cite{Mentzel2022fast} taking inspiration from the global gamma index~\cite{low1998technique}, used for clinical validation of treatment plans. It is defined as:
\begin{equation}
    \delta = \frac{D_{gen}-D_{MC}}{max(D_{MC})}\,,
\end{equation}
where $D_{gen}$ is the deposited energy predicted by the VAE, while $D_{MC}$ is the deposited energy obtained by the Monte Carlo simulation. 
As reconstruction performance measure we consider the 3\% passing rate, which is the percentage of voxels with a delta index smaller than 3\%.
On the test set our Network reaches $98.6 \pm 0.3 \%$ of voxels with 3\% passing rate.

\begin{figure}[h!]
\begin{center}
\includegraphics[width=.999\textwidth]{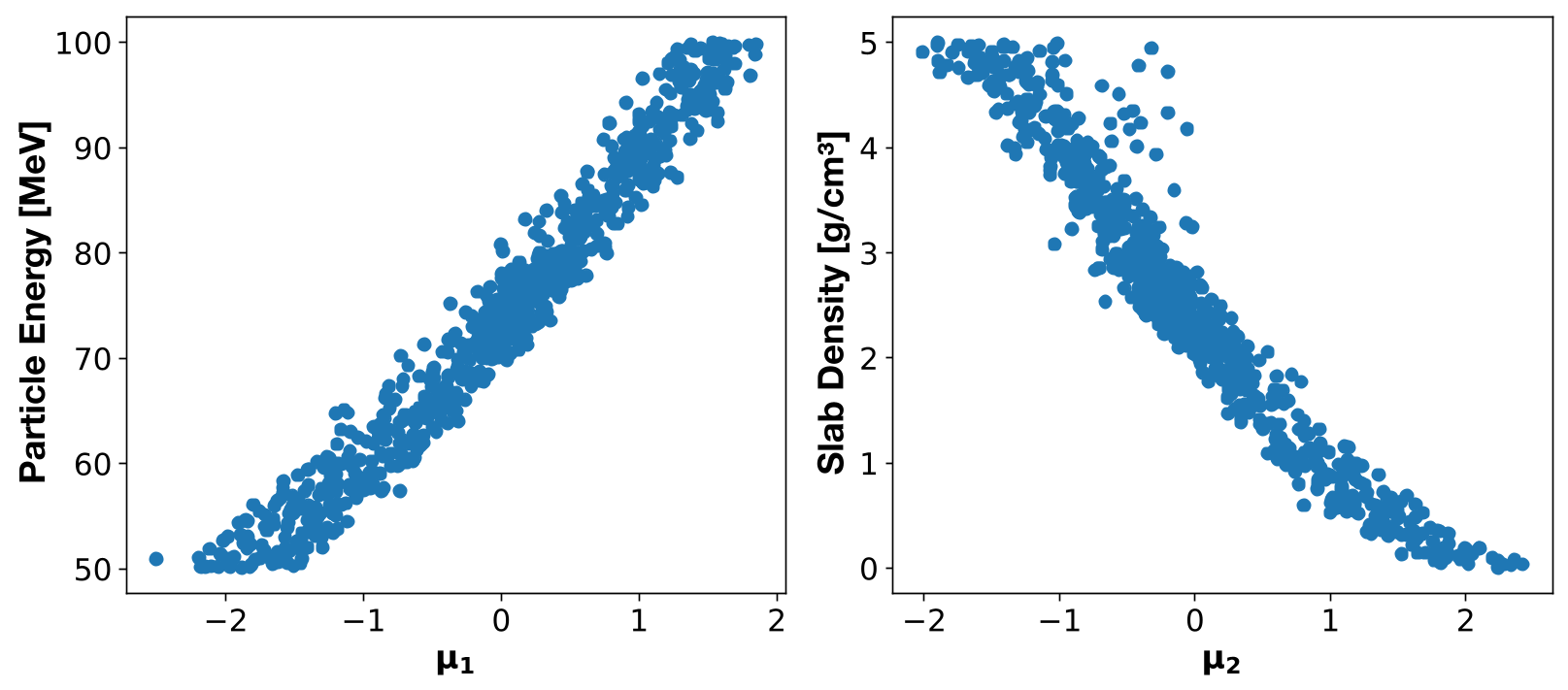}
\caption{\textbf{Correlation between physical parameters and latent space variables}. The latent variables $(\mu_1, \mu_2)$ are highly correlated with particle energy and slice density. Thanks to this relation it is possible to generate new samples conditioned to those parameters.}
\label{fig:dose_arsini:latent}
\end{center}
\end{figure}

\paragraph{Latent Space and generation}
Once the network is trained, it is possible to generate new samples of energy distributions by picking and decoding a point from the latent space.
Thus, to generate new samples conditioned to desired physical parameters, a relation between such parameters and the latent space variables is needed.
Looking at the latent space, we found that the network has well decoupled and reconstructed the information about those physical parameters. Indeed, after applying a 2-D rigid rotation, which, by definition, does not change the latent space structure, a strong correlation between the two physical parameters of the simulated data, \ie particle energy and slab density, and the two latent variables emerges. 
Such a result is shown in Fig.~\ref{fig:dose_arsini:latent}.

Thanks to these relationships, the generation of deposited energy distribution for a given particle energy and slice density is straightforward. 
Once the desired physical parameters are chosen, it is possible to link them to a point in the latent space using a fit and then, decode that point to obtain the corresponding deposited energy distribution.
The relation between latent variables and physical parameters presents some scattering, especially near the boundaries of the parameter space. This effect is due to the non perfect decoupling of the latent space in the whole parameters' range. The width of the scattered relations represents, indeed, the residual inter-correlations.
This effect on large parameters' range is expected for a VAE, but in fact we designed our model to generate samples in the limited parameters' range of the test set, where a quasi-perfect decoupling is possible.

In order to test the generative capabilities of the Network we performed a third degree polynomial fit on the relations between latent variables and parameters. Then we considered the physical parameters of the examples in our test and validation sets. Using the results of the fit, we converted physical parameters to latent variables and used the decoder to generate the corresponding samples. We finally computed the $\delta$ index between the original and the generated samples.
The 3\% passing rate on the test set reaches $98.4 \pm 0.5 \%$, a value that is compatible with the reconstruction result.

\paragraph{Discussion}
We presented a Graph VAE that can generate deposited energy distributions inside a voxelized material. 
There are two main advantages to our deposited energy distribution generation technique.
First, the described generation pipeline represents a way to write the deposited energy distribution in a volume as a continuous function of beam parameters and medium density.
Secondly, there is a huge gain from the point of view of computing time. Generating one example using Monte Carlo simulation with 4 threads and 10\,000 primaries on our CPU (HP Z2 Tower G5 Workstation) took around 82 seconds, while our network, on the same device, needs only 0.02 seconds, which corresponds to a speedup of a factor of $10^3-10^4$.
It is also worth considering that the deposited energy distributions for treatment planning are generally simulated with 1 million primaries, which make them even more computationally demanding. 
In contrast, the computing time for our network is independent of the number of primaries, because the DL model is trained to reproduce the per-particle dose distribution. In addition, the DL model's computing time can be further reduced on GPUs, thus the speed up with respect to Monte Carlo simulation is even greater.
Future works will be focused on testing the network on more complex materials and finally, on a patient's Computed Tomography (CT) scan. 

\subsection{Quantification of Uncertainties in a Model-Based Iterative Reconstruction Algorithm for Proton Computed Tomography}\label{sec:pct}

Proton computed tomography (pCT) is a novel medical imaging modality that provides the three-dimensional distribution of \emph{relative stopping power} (RSP) in the scanned object (``phantom''). As the RSP is the measure of how the proton beams deposit their energy in \eg a human body, RSP images form the basis of treatment planning procedures in proton radiation therapy. List-mode pCT establishes the RSP image from the energy losses of millions of individual protons, shot through the phantom on various paths. Several research groups are currently working on pCT detector hardware as well as simulation, tracking and tomographic reconstruction software~\cite{poludniowski_proton_2015,johnson_review_2018}. This contribution presents our work to quantify uncertainties in the tomographic reconstruction step of the Bergen-pCT system~\cite{alme_high-granularity_2020}.

\subsubsection{Proton CT and Model-Based Iterative Reconstruction}
In the setup of the Bergen pCT collaboration~\cite{alme_high-granularity_2020}, a digital tracking calorimeter (DTC) is used to measure the position $x_{i, \text{out}}$, direction $\dot x_{i, \text{out}}$, and energy of the individual protons $i=1,2,\dots,m$ after they left the phantom. The position $x_{i, \text{in}}$, direction $\dot x_{i, \text{in}}$ and energy before entering the object are statistical properties of the proton beam source. The energy difference before and behind the phantom can be converted to a \emph{water-equivalent path length} (WEPL) $w_i$.

Here, we substitute the measurement by a Monte Carlo simulation using GATE~\cite{GATE} and \geant~\cite{GEANT,allison_geant4_2006,allison_recent_2016} in a similar setup as in Section~\ref{sec:ad-of-geant4}, and merely focus on the list-mode tomographic reconstruction task given the proton positions, directions and energies. 

This task can be modelled as a least-squares problem in the following way. 
First, the unknown path of each proton through the phantom must be estimated from $x_{i,\text{in}}$, $\dot x_{i,\text{in}}$, $x_{i,\text{out}}$, and $\dot x_{i,\text{out}}$. The (extended) most likely path (MLP) formalism~\cite{schulte_maximum_2008,krah_comprehensive_2018} provides such estimations based on the model of multiple Coulomb scattering by Lynch and Dahl~\cite{lynch_approximations_1991} and Gottschalk~\cite{gottschalk_multiple_1993}. 

Then, given the estimated path of length $l_{i}$, the (measured or simulated) WEPL $w_i$ 
should satisfy the equation:
\begin{equation}\label{eq:integral-rsp-wepl}
    w_i = \int_0^{l_{i}} r(x_i(s)) \,\mathrm ds\,,
\end{equation}
where $x_i (s)$ is the position of the $i$-th proton along the path and $r(x_i(s))$ is the RSP at this position. For every proton $i=1,\dots,m$, an equation like~\eqref{eq:integral-rsp-wepl} is obtained in this way, with a known WEPL $w_i$ and an unknown (and sought) RSP distribution $r$. Discretizing $r$ as a 3D image with $n$~voxels,  we obtain a linear system of equations $w=A_{\text{x}} r$ with $A_{\text{x}} \in \mathbb{R}^{m\times n}$, where the matrix entry $(A_{\text{x}})_{ij}$ is related to the intersection length between the $i$-th proton path (estimated based on $x_{i,\text{in}}, \dot x_{i,\text{in}}, x_{i,\text{out}}, \dot x_{i,\text{out}}$) and the $j$-th voxel. Here, $\text{x}$ is the collection of $x_{i,\text{in}}, \dot x_{i,\text{in}}, x_{i,\text{out}}, \dot x_{i,\text{out}}$ for all protons $i=1,\ldots,m$. The linear system is sparse, overdetermined ($m>n$) and can only be solved in a least-squares sense. Typically, this is done by iterative algorithms like ART or DROP \cite{penfold_techniques_2015} on a GPU. 

\subsubsection{Model of Input Uncertainty}

Various types of uncertainties may arise during the pCT process and, thus, influence the process of the reconstruction of the 3-D image from proton positions and directions.  
As a long-term scope of this work, uncertainties shall be included in the process of designing a tracking calorimeter for pCT. For this purpose, we investigate on the effect of local perturbations of the input positions of protons by means of a differentiability analysis as well as on the effect of random inputs, focussing on proton positions, with the help of strategies for uncertainty propagation in a probabilistic description. 

Uncertainties caused by calibration errors, measurement errors, or errors in the track reconstruction process may affect the inputs $w$ and $\text{x}$ to the RSP reconstruction. It is then of interest to quantify their effect on the reconstructed RSP image $r$. 
In the following, we focus on a scenario with an uncertain but uniform shift $\Delta x_{\text{in}}$ of all proton positions $x_{i,\text{in}}$ in front of the DTC. Such a systematic measurement error may, \eg, arise from inaccuracies in the calibration of the magnets directing the proton beam and the positioning of the DTC, and lie in the range of a few millimeters. We model this error by a normally distributed random variable $\Delta X_{\text{in}}$ with mean $0$ and standard deviation given by $3\sigma = \SI{4}{\milli\meter}$. This is only a modeling assumption and not based on actual measurements, but it is based on the experience that variations in the range of $\pm \SI{4}{\milli\meter}$ are observed with a high probability ($99.7\%$ corresponding to the $3\sigma$-region of $\SI{4}{\milli\meter}$). 
Realizations $\Delta x_{\text{in}}$ of the random variable shift the proton positions, \ie, instead of the hit positions $x_{\added{i,}\text{in}}$ the DTC measures the perturbed positions $x_{\added{i,}\text{in}}+\Delta x_{\text{in}}$. 
To model more complex scenarios, like non-uniform random variations in the $x_{i,\text{in}}$ or the other measured quantities, a moderately larger number of random variables would have to be used.

\subsubsection{Analysis of Local Perturbations}
The image of a Gaussian distribution with mean $\mu$ and covariance matrix $\Sigma$ under an affine-linear function
\begin{equation}\label{eq:rsp-linear-function}
f: \mathbb{R}^q \rightarrow \mathbb{R}^{q'},~ z \mapsto J\cdot z + s
\end{equation}
is again a Gaussian distribution, with mean $f(\mu)$ and covariance matrix $J \cdot \Sigma \cdot J^T$. In our setup, the output $r$ depends on the uncertain shift $\Delta x_{\text{in}}$ in a non-linear way. If the dependency was sufficiently smooth, its linearization \eqref{eq:rsp-linear-function} could still be used to approximately quantify the uncertainty in $r$, by Taylor's theorem; $s$ would be the reconstructed image based on the measured $\added{\text{x}}$ and $w$, and $J$ would be the Jacobian matrix, which could be computed by AD/DP. However, our previous work \cite{aehle-derivatives-2023} indicated that the usual way of defining $(A_\text{x})_{ij}$ as a mean chord length if the $i$-th proton path and the $j$-th voxel intersect, and zero otherwise, leads to a piecewise differentiable dependency that achieves most of its evolution through jumps. Methods relying on linearization are not applicable under these circumstances. Further work might employ the ``fuzzy voxels'' approach~\cite{aehle-derivatives-2023} that led to a piecewise differentiable dependency whose evolution between the jumps better resembled the global behaviour, at the price of increased reconstruction time and blurrier solutions.

\subsubsection{Probabilistic Analysis}
Apart from a local analysis that is only valid for small perturbations, it is of interest to observe and quantify the effects of uncertainties in a probabilistic setting. This may also help to analyze modeling strategies that try to include uncertain inputs. 

There exist different methods to propagate input uncertainties in modeling. As already described above, we may express calibration errors using a moderate number of random variables $Z$ with realizations $z_i$. This makes the use of projection methods like non-intrusive discrete projection computationally much more affordable than sampling-based strategies like classical Monte Carlo sampling. In the following, we consider a non-intrusive polynomial chaos approach (see \eg Ref.~\cite{Xiu2007}), which is also referred to as pseudo-spectral approach. 
In this approach, the probabilistic outcome $q$ is expanded in terms of polynomials $\phi_i$ that are orthogonal with respect to the probability density function $\rho_Z$ of the input random variables $Z$ (compare with Ref.~\cite{ghanem1991}). For a one-dimensional random input $Z$ results in the expression:
\begin{align}
    f(Z)= \sum\limits_{i=0}^{\infty} \hat{f}_i\phi_i(Z)\,,
\end{align}
with deterministic coefficients $\hat{f}_i=\gamma_i^{-1} \int\limits_{\mathbb{R}} f(z) \phi_i(z) \rho_Z(z)\,\mathrm dz$ and $\int\limits_{\mathbb{R}} \phi_i(z) \phi_j(z) \rho_Z(z)\,\mathrm dz = \gamma_i \delta_{ij}$. When applied to find statistical quantities or to sample from this distribution, the infinite expansion is truncated.
In the non-intrusive approach, the coefficients are approximated by employing a quadrature rule that is suitable for the used polynomials, \eg Hermite polynomials for normally distributed variables. The computational effort of the quadrature can be reduced by using sparse grid quadrature rules.

\begin{figure}
    \centering
    \includegraphics[height=0.18\textheight]{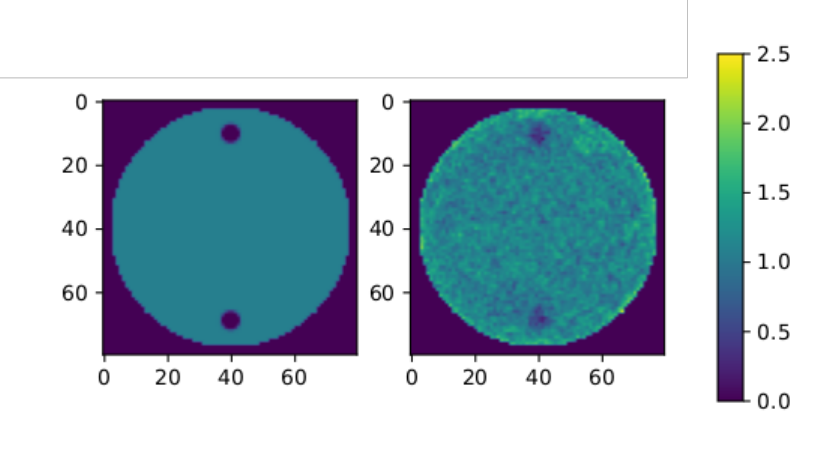}
    \caption{RSP values for a slice ($80\times 80$ voxels, each sliced voxel is $\SI{2}{\milli\meter} \times \SI{2}{\milli\meter}$) of the approximated original phantom (left) and a reconstructed image (right) with circular air holes in the upper and lower part. Note that the original phantom has more holes with less contrast.}
    \label{fig:phantom}
\end{figure}

For this study, we measure the quality of the reconstruction by observing the reconstruction error of the RSP values $r_{\text{recon}}$ in the voxels, \ie:
\begin{align}
    \Delta r:= ||r_{\text{recon}}-r_{\text{org}}||_2^2\,,
\end{align}
with $r_{\text{org}}$ being a approximation of the original RSP representation of the phantom used in the simulation with GATE. The aim is to observe the influence of uncertainties on this reconstruction error. We restrict the error calculation to regions of interest for a specific test case. For this case we have two circular air holes in a phantom (see Fig.~\ref{fig:phantom}).

Uncertainties in the measurements may be specifically addressed in modeling when using an extended most likely path formalism~\cite{krah_comprehensive_2018}.
Here, we provide the corresponding covariance matrices alongside $x_{\added{i,}\text{in}}, \dot{x}_{\added{i,}\text{in}}$ and $x_{\added{i,}\text{out}}, \dot{x}_{\added{i,}\text{out}}$ and extend the expression for the maximum likelihood estimate. 
With the strategies of Monte Carlo sampling (MC) and the non-intrusive polynomial chaos method (PC) we may compare the effects on the quality of the reconstruction for both path estimation formalism, the original MLP~\cite{schulte_maximum_2008} and the extended MLP (EMLP)~\cite{krah_comprehensive_2018}. The resulting expected reconstruction quality and its variance are shown for both formalisms in Table~\ref{tab:tab_mc_pc}.

\begin{table}
    \centering
    \begin{tabular}{ |c|c|c|c|c|c| }
 \hline
 formalism & method & points & $\Delta r(E{\Delta x})$ & $E(\Delta r)$ & Var$(\Delta r)$\\ \hline
MLP & MC & 2000 & 0.0047410 & 0.0047158& 3.1852e-08\\ 
MLP & PC & 60 & 0.0047410 & 0.0047251& 3.7064e-08\\ 
EMLP & MC & 2000 & 0.0039905 & 0.0039281 & 6.3499e-09\\ 
EMLP & PC & 60 & 0.0039905 & 0.0039128 & 6.9032e-09\\ 
 \hline
\end{tabular}
    \caption{Expected value $E$ and variance $Var$ of the reconstruction error for the different formalisms.}
    \label{tab:tab_mc_pc}
\end{table}

In general, we observe that the non-intrusive polynomial chaos approach will provide a good approximation of the expected value and the variance with less function evaluations (60) than the Monte Carlo method (2\,000). A more rigorous convergence analysis shows that, as expected, the polynomial chaos approach will converge faster than the Monte Carlo approach. However, we do not observe exponential convergence. This can usually only be observed for sufficiently smooth function which is not the case for the reconstruction error. 

\begin{figure}
    \centering
    \includegraphics{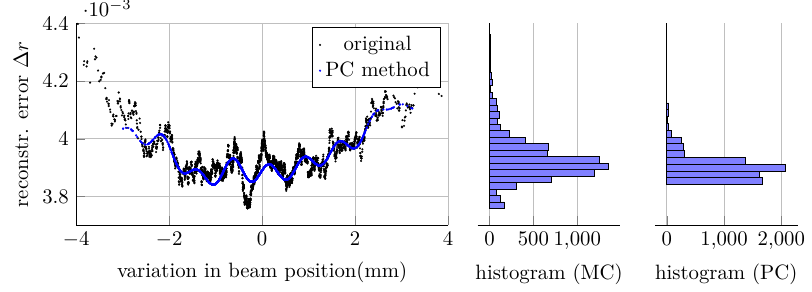}
    \caption{Sampled the distribution of the reconstruction error (left) for the original distribution (Monte Carlo sampling, black) and the \added{PC} expansion (blue) and corresponding histograms (right) sharing the $y$-axis with the left plot (reconstruction error).
    }
    \label{fig:comparison}
\end{figure}

In Fig.~\ref{fig:comparison}, which shows the sampled reconstruction error (black), we observe a high amount of fluctuations in the function. The sampled polynomial chaos expansion (blue) is smoother and it may only reflect a global behaviour. This can also be observed in the corresponding histograms. The general trend for observations with a high probability is similar. However, the values of a low reconstruction error around 0.0038 are not observed for the expansion. This shows that the expansion can be used to analyze average behaviour and outcomes with high probabilities (which is of interest for detector design), but it is not feasible for analyzing, \eg, outliers.  

\subsubsection{Conclusion and Outlook}
In this work, we have considered the effects of uncertainties on the image reconstruction in pCT using different modeling approaches. 
Regarding local perturbations, we observed that the algorithmic description of the reconstruction process is only piecewise differentiable. Jumps arise from the discrete computation of the set of voxels intersected by a proton path.  
For propagating the input uncertainty in the calibration, we could reduce the computational costs by using a non-intrusive discrete projection method. The modeling and propagation of uncertainties allowed us to assess the overall impact of the extended most-likely path approach on the quality of the reconstruction. Quantities like the expected value or the variance could be approximated with satisfying quality. In future work, it is intended to include the quantification of uncertainties in the design process of the calorimeter to find solutions that are more robust under perturbations. 

\clearpage
\section{Conclusions and Future Prospects}
\label{sec:concl}
The end-to-end optimization of detectors for fundamental physics measurements constitutes a class of challenging and multi-faceted problems whose solution requires the development of specialized models performing a number of unique tasks. So, \eg, an experiment tasked with detecting high-energy cosmic rays and an experiment built to study particle collisions have very little in common, despite the fact that the underlying physics responsible for the data-generating processes is the same.

This heterogeneity has until recently discouraged a synergistic effort in the direction of providing a general solution. Nevertheless, the common traits these problems share, and the large potential value of the optimization of instruments whose price tag ranges in the several million to several hundred million euros, motivated us to invest our research time in this area. Our goal is to create a library of solutions and a toolkit that may empower the tackling of still harder use cases, by considering problems of comparatively small or moderate complexity, while still already at the edge of our capability.

Following the above plan, this document collects discussions of a range of use cases on which we are progressing toward end-to-end modeling and optimization solutions for experiments and instruments in fundamental physics. As a work in progress, each individual project discussed here only offers a snapshot of the current state of the art, rather than final solutions or crystallized results which, for the considered use cases, may or may not be far behind.

\section*{Acknowledgements}
{\tolerance=3000 This work was supported by the U.S. Department of Energy, under DOE Contract No.~\texttt{DE-AC02-76SF00515}, the Office of Science, Office of Basic Energy Sciences. This work was partially supported by the EU Horizon 2020 Research and Innovation Programme under grant agreement No.~\texttt{101021812} ("SilentBorder"). Some of the research leading to these results has received funding from the Basic Research Program at the National Research University Higher School of Economics. Max Aehle gratefully acknowledges funding from the research training group SIVERT by the German federal state of Rhineland-Palatinate. Max Aehle, Lisa Kusch and Nicolas R.\ Gauger contributed on behalf of the Bergen pCT collaboration, whose current members include: Johan Alme, Gergely Gábor Barnaföldi, Tea Bodova, Vyacheslav Borshchov, Anthony van den Brink, Viljar Eikeland, Gregory Feofilov, Christoph Garth, Nicolas R.\ Gauger, Ola Grøttvik, Håvard Helstrup, Sergey Igolkin, Ralf Keidel, Chinorat Kobdaj, Tobias Kortus, Viktor Leonhardt, Shruti Mehendale, Raju Ningappa Mulawade, Odd Harald Odland, George O'Neill, Gábor Papp, Thomas Peitzmann, Helge Egil Seime Pettersen, Pierluigi Piersimoni, Maksym Protsenko, Max Rauch, Attiq Ur Rehman, Matthias Richter, Dieter Röhrich, Joshua Santana, Alexander Schilling, Joao Seco, Arnon Songmoolnak, Ganesh Tambave, Ihor Tymchuk, Kjetil Ullaland, Monika Varga-Kofarago, Lennart Volz, Boris Wagner, Steffen Wendzel, Alexander Wiebel, RenZheng Xiao, Shiming Yang, Hiroki Yokoyama, Sebastian Zillien. Jennet Dickinson (also supported by the DOE Early Career Research Program) and Lindsey Grey are supported by Fermi Research Alliance, LLC under Contract No.~\texttt{DE-AC02-07CH11359} with the Department of Energy (DOE), Office of Science, Office of High Energy Physics.
Borja S. González is grateful for the financial support by the FCT PhD grant PRT/BD/151553/2021 under the IDPASC program.
Fredrik Sandin's work was supported by the Jubilee and Kempe Foundations, project No.~\texttt{JCSMK JF-2303}.
Pietro Vischia’s work was supported by the MINECO and the EU through the Ramón y Cajal program under the Project No.~\texttt{RYC2021-033305-I}.\par} 

\bibliographystyle{unsrturl}
\bibliography{main.bib}

\end{document}

%% file: lstinit.tex
\makeatletter
\let\old@lstKV@SwitchCases\lstKV@SwitchCases
\def\lstKV@SwitchCases#1#2#3{}
\makeatother 
\usepackage{lstlinebgrd}
\makeatletter
\let\lstKV@SwitchCases\old@lstKV@SwitchCases

\lst@Key{numbers}{none}{%
    \def\lst@PlaceNumber{\lst@linebgrd}%
    \lstKV@SwitchCases{#1}%
    {none:\\%
     left:\def\lst@PlaceNumber{\llap{\normalfont
                \lst@numberstyle{\thelstnumber}\kern\lst@numbersep}\lst@linebgrd}\\%
     right:\def\lst@PlaceNumber{\rlap{\normalfont
                \kern\linewidth \kern\lst@numbersep
                \lst@numberstyle{\thelstnumber}}\lst@linebgrd}%
    }{\PackageError{Listings}{Numbers #1 unknown}\@ehc}}
\makeatother